\documentclass[twocolumn]{aastex62}

\renewcommand{\edit}[1]{\textcolor{red}{#1}}

\def\nustar{{\it NuSTAR\/}}
\def\beppo{{\it Beppo-SAX\/}}
\def\RXTE{{\it RXTE\/}}
\def\xmm{{\it XMM-Newton\/}}
\def\chandra{{\it Chandra\/}}
\def\flux{erg~s$^{-1}$~cm$^{-2}$}
\graphicspath{{./}{figures/}}

\received{September 27, 2020}
\revised{October 27, 2020}
\accepted{November 17, 2020}
\shorttitle{NuSTAR Observations of Abell 2163}
\shortauthors{Rojas et al.}
\begin{document}

\title{\nustar\ Observations of Abell 2163: Constraints on Non-thermal Emission} %\footnote{Released on January, 8th, 2018}}

\author [0000-0002-8882-6426]{Randall A. Rojas Bolivar}
\affil{Department of Physics \& Astronomy,
University of Utah,
115 South 1400 East, Salt Lake City, UT 84112, USA}

\author[0000-0001-9110-2245]{Daniel R. Wik}
\affil{Department of Physics \& Astronomy,
University of Utah,
115 South 1400 East, Salt Lake City, UT 84112, USA}

\author[0000-0002-1634-9886]{Simona Giacintucci}
\affiliation{Naval Research Laboratory, 
4555 Overlook Avenue SW, Code 7213, 
Washington, DC 20375, USA}

\author[0000-0002-9112-0184]{Fabio Gastaldello}
\affil{Istituto di Astrofisica e Fiseca Cosmica-Milano
Via Edoardo Bassini, 15
20133 Milano MI, Italy}

\author[0000-0002-3363-0936]{Allan Hornstrup}
\affil{DTU Space, Technical University of Denmark,
Elektrovej Building 327, DK-2800
Kgs Lyngby, Denmark}

\author[0000-0001-5839-8590]{Niels-Jorgen Westergaard}
\affil{DTU Space, Technical University of Denmark,
Elektrovej Building 327, DK-2800
Kgs Lyngby, Denmark}

\author[0000-0002-2114-5626]{Grzegorz Madejski}
\affil{SLAC National Accelerator Laboratory
2575 Sand Hill Rd
Menlo Park, CA 94025}

%\collaboration{(AAS Journals Data Scientists collaboration)}




% DRW
% The "\nustar" command does not leave a space after the word, so one needs to be added
% manually with "\ ", e.g. "The \nustar\ observatory...".  However, if you don't need a space,
% then the trailing forward-slash isn't needed, e.g., "with \nustar, the".

\begin{abstract}
Since the first non-thermal reports of inverse Compton (IC) emission  from the intracluster medium (ICM) of galaxy clusters at hard X-ray energies, we have yet to unambiguously confirm IC emission in observations with newer facilities. \RXTE~detected IC emission in one of the hottest known clusters, Abell 2163 (A2163), a massive merging cluster with a giant radio halo---the presumed source of relativistic electrons IC scattering CMB photons to X-ray energies. The cluster's redshift ($z\sim0.2$) allows its thermal and non-thermal radio emission to fit \nustar's FOV, permitting a deep observation capable of confirming or ruling out the \RXTE~report. The IC flux provides constraints on the average magnetic field strength in a cluster. To determine the global diffuse IC emission in A2163, we fit its global \nustar\ spectrum with four models: single (1T) and two-temperature (2T), 1T+power law component (T$+$IC), and multi-temperature+power law (9T$+$IC).  Each represent different characterizations of the thermal ICM emission, with power law components added to represent IC emission. We find the 3--30 keV spectrum can be described by purely thermal emission, with a global average temperature of $kT=(11.8\pm0.2)$ keV. The IC flux is constrained to
$<$4.0$\times$$10^{-12}$~erg~s$^{-1}$~cm$^{-2}$ using the 1T$+$IC model and $<$1.6$\times$$10^{-12}$~erg~s$^{-1}$~cm$^{-2}$ with the more physical 9T$+$IC model, both to 90$\%$ confidence levels. Combining these limits with 1.4 GHz diffuse radio data from the VLA, we find the average magnetic field strength to be $>$$0.22 \mu$G and $>$$0.35 \mu$G, respectively, providing the strongest constraints on these values in A2163 to date.
%\todo{It is poor practice to not make labels for figures (inside the caption text at the end, add {\tt \textbackslash label\{fig:labelname\}} and wherever you reference a figure by name, use ``Figure$\sim${\tt \textbackslash ref\{fig:labelname\}}".}

\end{abstract}


%% sections are demarcated by \section and \subsection, respectively.

\section{Introduction}
\label{sec:intro}

%DRW: why give size?  what determines the smallest size?  clusters are better described by their mass --> change clause to "reaching masses of up to $10^{16}~$M$_\odot$."
    Galaxy clusters are the largest form of gravitationally bound objects known in the Universe, reaching masses of up to $\sim$$10^{15}~M_\odot$.
%DRW:
Due to the presence of a deep gravitational potential well, the gas or intracluster medium (ICM) heats up to temperatures between $10^{7-8}$~K, or equivalently,
% technically units of keV are not temperatures, but energies - better to use Kelvin units first, then express as $kT \sim 10$~keV.
$kT \sim 1$--10 keV, which provides pressure support to the gas against gravitational collapse.
%cite recent work using the XCOP sample too, Ettori 2018 I think
The temperature and density of the cluster can be used to calculate the pressure profile, which in turn can be used to determine the mass under the assumption of hydrostatic equilibrium \citep[e.g.,][]{1993ApJ...407L..49B, 2009astro2010S.305V, 2019A&A...621A..39E}.

%(Bahcall et al. 1993, Vikhlinin et al 2009).
%This pressure can be used to determine the temperature profile of the galaxy cluster, which can then be used to calculate the mass of the cluster.
% for cosmology, we don't care about the mass profile, just the total mass of one cluster -- what gives us cosmological parameters is the number of clusters per unit mass
%One can use clusters to provide estimates on the amounts of baryonic matter, dark matter, and dark energy present in the Universe in order to constrain cosmological parameters.

% something like: "When two clusters merge, the gas eventually gets heated to the new virial temperature determined by their combined mass."
When two clusters undergo a merger event, the gas is heated by shock fronts and mixed by turbulence, bringing it to a new virial temperature. Merging clusters also host diffuse radio synchrotron emission (radio halos and relics) that require a non-thermal electron population presumably made visible by some kind of Fermi-like acceleration process related to the merger \citep{1999NewA....4..141G}. Surveys by \citet{2016MNRAS.457.4515R} show that nearly one half of all known galaxy clusters are undergoing a merger event. The same accelerated relativistic electrons radiating synchrotron emission in the radio must produce non-thermal emission in the form of inverse Compton (IC) scattering through interactions with cosmic microwave background (CMB) photons.
%reheating of the gas causes the relativistic particles in the intracluster medium to gain energy by means of the
% first order is in shocks and we think produces radio relics, while radio halos are thought to be produced by accel. caused by turbulence - 2nd order Fermi accel.
%first order Fermi acceleration. Due to this
% first mention of "re"acceleration - need to first say that b/c shocks are weak (have low Mach numbers), the accel. efficiency is low, so particles can't be accel. directly from the thermal pool - there needs to be pre-existing population of mildly rel. particles already that gain enough energy by these weak shocks and turb. to become visible
%reacceleration, one would expect to see non-thermal emission in the form of Inverse Compton (IC) scattering present in observations.

    %Paragraph about ABELL 2163, previous temperature measurements, and previous IC scattering measurements in this cluster along with other ones (i.e. Bullet Cluster).

   Abell~2163 (hereafter, A2163) is one of the most massive clusters observed in the Universe
% maybe say "one of the more massive clusters..." - not sure how far it is from the top.  I think it is the hottest of the Abell clusters, although that's not a very scientific statement (but you can use it)
\citep{1994ApJ...436L..71M}. %(Markevitch et al. 1994). 
It is also one of the most distant and rich clusters within the Abell catalogue \citep{1958ApJS....3..211A, 1996ApJ...456..437M}. %(Markevitch et al. 1996). 
The cluster is currently undergoing a merger event \citep{2008A&A...481..593M}. %(Maurogordato et al. 2008). 
Original temperature measurements suggested a global temperature of 15 keV \citep{1992eocm.rept..177A}. %(Arnaud et al. 1992) \citet{2017ApJ...848...80M}. 
Past X-ray surveys have shown the gas distribution within the cluster to be non-isothermal, including an evident temperature gradient within the core of the cluster \citep{1994ApJ...436L..71M, 1996ApJ...456..437M}. %(Markevitch et al. 1994; Markevitch 1996; Markevitch) \& Vikhlinin 2001).
% phrase as "More recent studies suggest a lower temperature around 12-13 keV."
In later studies, the global temperature of the cluster has fallen to values between 12 keV and 13 keV \citep{2002ApJ...573L..69H}. % (Markevitch et al. 1994, Hansen et al. 2002). 

Claims have been made of non-thermal emission due to the inverse Compton scattering of CMB photons by the relativistic electrons present in the ICM within the cluster \citep{2001A&A...373..106F, 2006ApJ...649..673R, 2009MNRAS.399.1307M, 2014A&A...562A..60O}. 
Data taken by \beppo\ failed to detect significant IC emission, yielding an upper limit on the flux of the IC scattering ($F_{NT}$) between 20-80 keV at 90\% confidence level less than 5.6 $\times$ $10^{-12}$~\flux~\citep{2001A&A...373..106F}. A long observation \RXTE\ was argued to show a detection with large uncertainties of $F_{NT}$ $\sim$ $1.1^{+1.7}_{-0.7}$  $\times$ $10^{-11}$~\flux~\citep{2006ApJ...649..673R}, nearly 25\% of 
the integrated 3-50 keV flux, inconsistent with the \beppo\ upper limit. Using \chandra, \citet{2009MNRAS.399.1307M} argue a detection of $F_{NT}$ $\sim$ $3.9^{+1.0}_{-1.0}$  $\times$ $10^{-12}$~\flux~between 0.6-7.0 keV, which accounts for roughly 10\% of the integrated flux. 
A later 90\% upper limit of detection, $F_{NT}$ $<$ 1.2  $\times$ $10^{-11}$~\flux~\citep{2014A&A...562A..60O} %(Ota et al. 2014) 
was provided by work done using the Hard X-Ray Detector (HXD)  % (Takahashi et al. 2007) 
on the Suzaku satellite %(Mitsuda et al. 2007) 
in combination with XMM-Newton data.\\
\indent Using IC upper limits in conjunction with radio data yields a lower limit on the magnetic field \citep{1979ApJ...227..364R}. It is important to know the magnetic field strength of galaxy clusters because they play a role in energy distribution in the gas as well as contribute dynamically to physical processes occurring within clusters \citep{2002ARA&A..40..319C}. %(Carilli \& Taylor 2002). 
    Due to the large scale of clusters, knowing constraints on the magnetic field can also provide insight into cosmological magnetic fields and their evolution \citep{2018Galax...6..142V}. %(Vacca et al. 2018).

% also want to describe how radio detections plus IC upper limits yield B field lower limits, and why that's important

% for this paragraph, just say "In this paper, we present a deep NuSTAR observation of A2163 in order to detect or constrain IC emission.  In section 2, we discuss the observations and basic processing; in section 3, ... etc.  Unless otherwise stated, all errors are quoted at the 90% level and we assume a cosmology of ..., giving a distance of X to the cluster."
    %Paragraph about what we do with NuSTAR data
    %Using data obtained during a %10 ks!!! observation from NuSTAR (Nuclear Spectroscopic Telescope Array), we establish an upper limit on the IC scattering detections present within the ABELL 2163 cluster. By doing so, we can also obtain a lower limit on the magnetic field strength. We use radio data from the VLA (Very Large Array) in the 1.4 GHz range to calculate the amount of synchrotron radiation. Doing the ratio of the IC and synchrotron provides a lower limit on the magnetic field.

    In this paper, we present a deep \nustar\ (Nuclear Spectroscopic Telescope Array) \citep{2013ApJ...770..103H} observation of A2163 in order to detect or constrain IC emission. We also use data obtained from the VLA (Very Large Array) in the 1.4 GHz range to get constraints on the lower limit of the magnetic field strength. In Section~\ref{sec:observe}, we discuss the observations and how the data was processed. In Section~\ref{sec:background}, we show how the background was modeled as well as discuss the individual components within this model. In Section~\ref{sec:analysis}, we describe our analytical approach on the data and which model provides the best constraints. In Section~\ref{sec:Summary}, we summarize our findings with respect to previous studies and discuss future possibilities. We also provide Appendix~\ref{app:ARF}) containing information regarding an issue with this observation concerning auxiliary response file generation.
    %occurring with the getspecext.py program in {\tt nuproducts}. %maybe? no -- getspecext.py is not a nuproducts routine
% do we ever need to assume a cosmology?  only relevant if we quote luminosities, as they depend on the cluster's distance
For this paper, all errors are quoted at the 90\% confidence level. %and we assume a flat cosmology with $\Omega_{\rm M}$ = 0.23 and $H_{0} = 70$~km~s$^{-1}$~Mpc$^{-1}$, yielding a distance for the cluster of \edit{what??} at its redshift $z = 0.203$.

\section{Observations and Data Reduction} \label{sec:observe}

\subsection{X-ray} \label{sec:observe:xray}
\begin{figure}[h]
\centering
\includegraphics[scale=0.3]{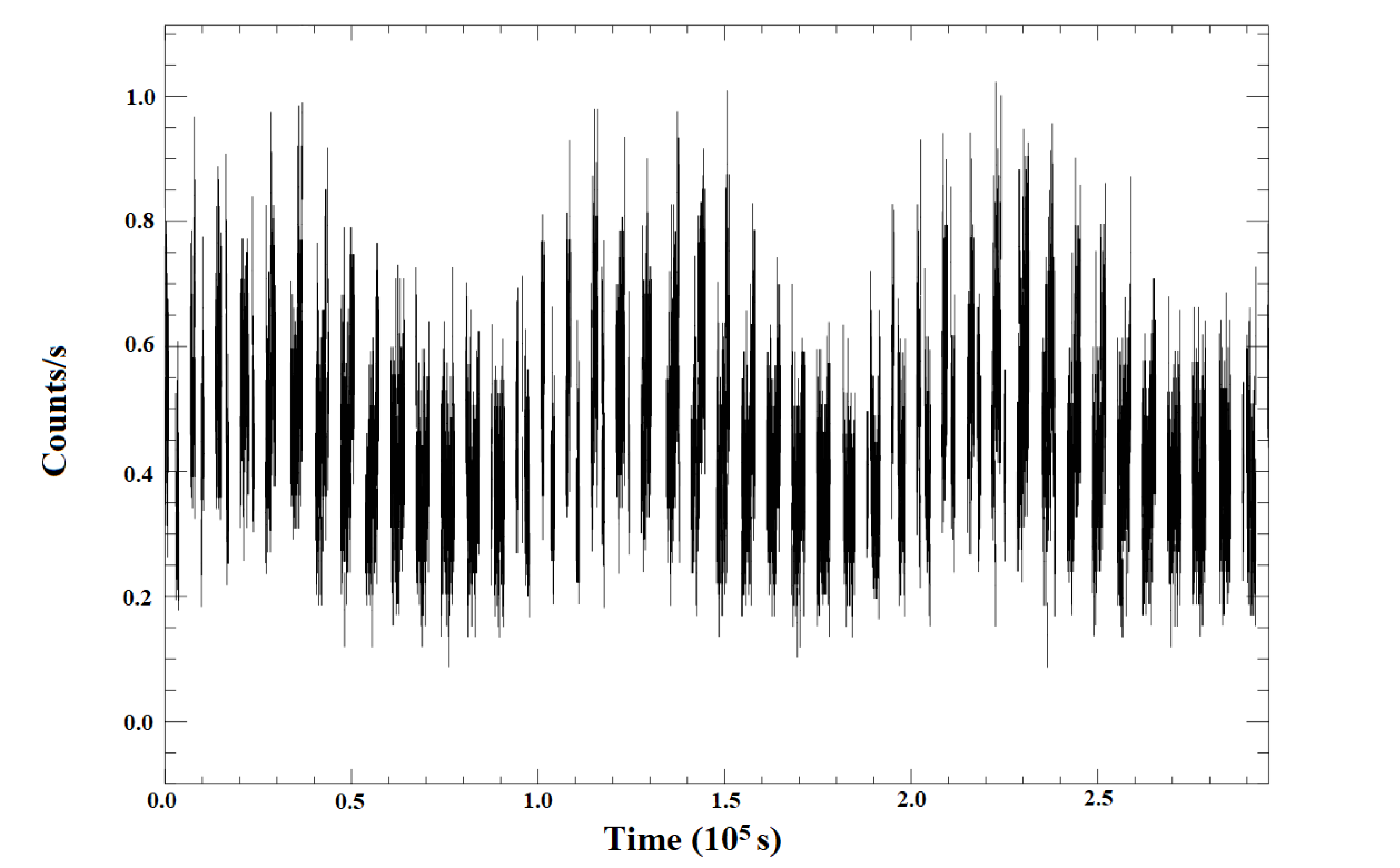}
\includegraphics[scale=0.3]{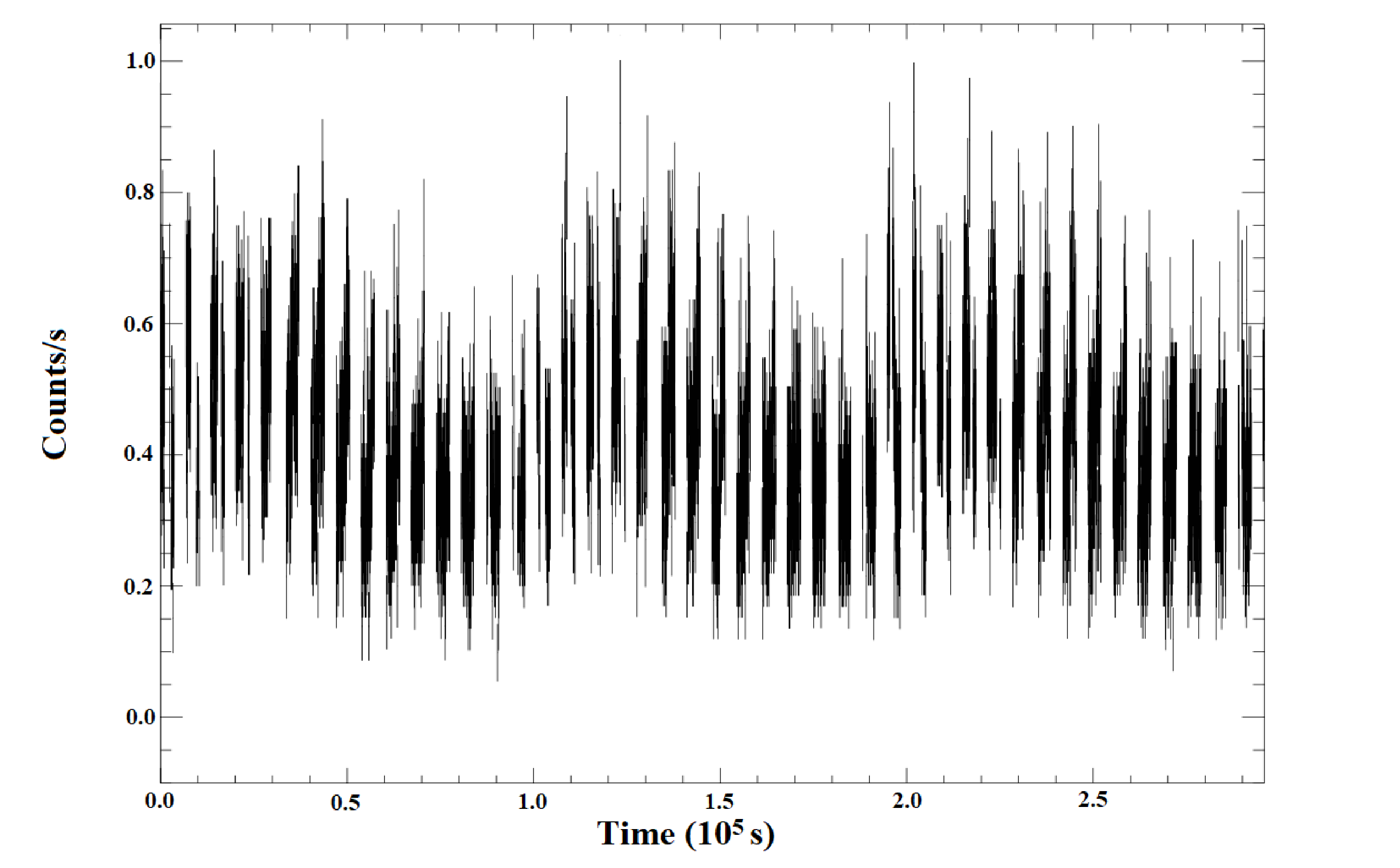}
\caption{
Filtered light curves for both the A (top) and B (bottom) telescopes following the process described in Section~\ref{sec:observe:xray}. The light curves have been filtered in the 50-160 keV energy range to eliminate SAA background contributions as well as the 1.6-20 keV range to remove solar activity background contributions.  
\label{fig:lightcurve}
}
\end{figure}

 A2163 was observed by \nustar\ for a net raw exposure time of 115 ks, including periods when the cluster was occulted by the Earth. The observation was carried out between March 23rd, 2016 and March 26th, 2016. To filter the data, we used standard pipeline processing from {\tt HEASoft} version~6.22 and {\tt NuSTARDAS} version~1.7.1. In order to obtain clean spectra for analysis, we needed to remove high background periods from the data since our measurement is most sensitive in the background-dominated regime. Normally, this is done by setting the {\tt STRICT} and {\tt TENTACLE} flags within the {\tt nupipeline} processing. The {\tt STRICT} mode identifies when the telescope passes through the South Atlantic Anomaly (SAA) by detecting high-gain shield single rates stored in the file for the SAA calculation algorithm. The {\tt TENTACLE} flag detects time intervals in which the CZT detectors have an increase in event count rates when crossing the SAA.\footnote{https://heasarc.gsfc.nasa.gov/docs/nustar/analysis/nustar\_swguide.pdf} %(Perri et al. 2017) Figure out how to site this, not on ADS.
% footnote is correct in this case
This filtering, however, can at times be too strict and remove good periods as well as miss high background periods. To combat this, we instead process the data with these flags turned off and derive a custom set of good time intervals (GTIs) manually. We create light curves using all the data from the instrument for the A and B telescopes separately by running the {\tt lcfilter} command. These light curves are then binned in bins of 100~s. We manually identify and exclude time intervals where the count rate is above the local distribution of rates, which is carried out in three passes. The first pass is to eliminate high background periods due to the SAA at harder energies (50-160 keV). The second pass is again performed in the same energy range to further reduce smaller contributions from the SAA after the larger background intervals have been removed. The final pass is done in low energy ranges (1.6-20 keV) to remove high background periods due to solar activity. This process reduced the exposure time to 112 ks in both the A and B telescopes. After filtering the data based on our new set of good time intervals (GTIs), there were no notable fluctuations in the final light curve, suggesting a clean background (see Figure~\ref{fig:lightcurve}).

%\todo{good chance the referee will want to see this cleaned light curve - consider making such a figure (perhaps while waiting for feedback from coauthors)}

% better to phrase this way: first, state that we need to remove periods when the background is high, since our measurement is sensitive to the background level.  this is usually achieved by setting the "STRICT" and "TENTACLE" flags during the nupipeline processing (and explain what it does).  however, this filtering can remove good periods as well as miss high background periods, so instead we process the data without these flags and derive a custom set of good time intervals (GTIs) manually.  we create light curves, binned by 100s, and interactively exclude times where the rate is above the local distribution of rates in 3 passes (explaining why 3 times).

%No strict filtering criteria were applied to account for the South Atlantic Anomaly and the increase in activity located within a tentacle shaped region nearby. Light curves were filtered manually three separate times (go back and check what was different every time)
    %,reducing exposure time to (insert final exposure time).

    Using the new GTIs, we reprocessed the data with {\tt nupipeline}. The cleaned files were used to generate images with {\tt XSELECT}. Exposure maps were created using {\tt nuexpomap}. For spectral fitting, spectra, response matrices, and auxiliary response files (RMFs and ARFs, respectively) were initially made with {\tt nuproducts}. Generally, {\tt nuproducts} has a flag ({\tt extended = yes}) that weights the ARF based on the distribution of events within the chosen extraction region. This is under the
% a little imprecise - it assumes the photons reflect the true underlying distribution incident on the telescope, which is not true since nustar's psf scatters them; instead of "actual extent", something like "... assumption that the observed spatial distribution of source photons is identical to the incident distribution of the source."
assumption that the observed spatial distribution of source photons is identical to the incident distribution of the source, which in principle is not entirely true. It appears that in the case of this particular observation, {\tt nuproducts} produced ARFS are normalized using an incorrect weighting scheme.
%ha, wish we could leave that in!  just to be clear, the offset isn't in the optical axis, but where the software thinks the optical axis is

%Due to an imprecision in the way the software finds the optical axis, something goes wonky in the weighting and gives incorrect ARFs %(super scientific sentence!).
%(This issue was not present when generating point source ARFs,
% say something like "and seems to be related to how arfs are combined when generating extended arfs, assuming an average placement of the optical axis that, in this particular case, offset from the true position"
%and seems to not be an issue in other observations using the same methods that we know of). This appears to be related to how the ARFs are combined when generating the extended ARFs, assuming an average placement of the optical axis, that in our case is being offset from the true position.
% actually, it just replicates what nuproducts should be doing by deriving the actual location of the optical axis
To account for this issue, we calculate ARFs for a given region following the intended procedure of {\tt nuproducts, extended = yes}, with a grid of point source effective area curves across the region summed and weighted by the relative number of net counts in that vicinity based on a background-subtracted image. We verify that this procedure reproduces similar results as point source ARFs created by {\tt nuproducts}
(see Appendix~\ref{app:ARF} for details). We present cleaned images of the cluster, along with the regions used in this analysis, in Figure~\ref{fig:region}.
%and applied it to extended sources to obtain new files (see Appendix I for more detailed description). %(Maybe ARF plots I made go here, or in Appendix?)
%\begin{figure*}
%\includegraphics[scale=0.6]{arfsnoext.PNG}
%\includegraphics[scale=0.6]{extendedarfs.PNG}
%\includegraphics[scale=0.6]{workingextarfs.PNG}
%\caption{\footnotesize{Top Left: A plot showing the point source ARF, with the green line representing the normalized ARF obtained from a region located at the optical angle. As expected, ARFs grabbed from regions surrounding this one should fall below the line. Top Right: ARFs generated using {\tt nuproducts} show that the optical axis is not being properly located, as ARFs generated from the same regions as before now lie above the normalized ARF. Bottom Left: ARFs generated using our own custom script with proper optical axis location fixes the issue.}}
%\end{figure*}

\subsection{Radio} \label{sec:observe:radio}

We reanalyzed archival VLA data of A\,2163 at 1.4 GHz (project AF328, same data used in \citet{2001A&A...373..106F}).
The cluster was observed in C-array configuration for approximately 4 hours and twice in D configuration for a
total of 3.4 hours. All observations used two intermediate frequency (IF) channels centered at 1365 MHz and
1435 MHz, with a bandwidth of 50 MHz/IF for the D configuration observations and 25 MHz/IF in C configuration.

\begin{figure}[h]
\includegraphics[scale=0.4]{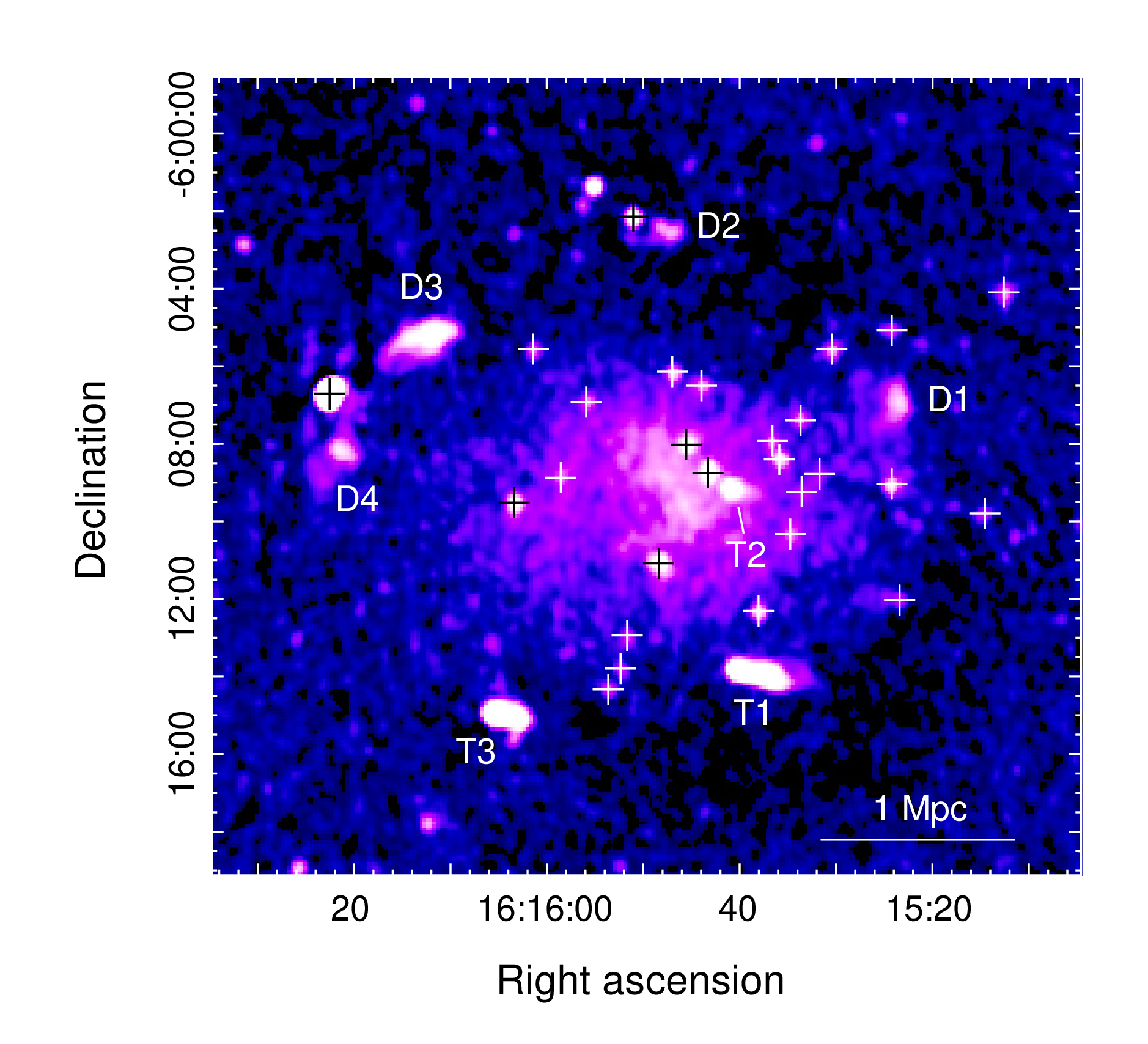}
\includegraphics[scale=0.4]{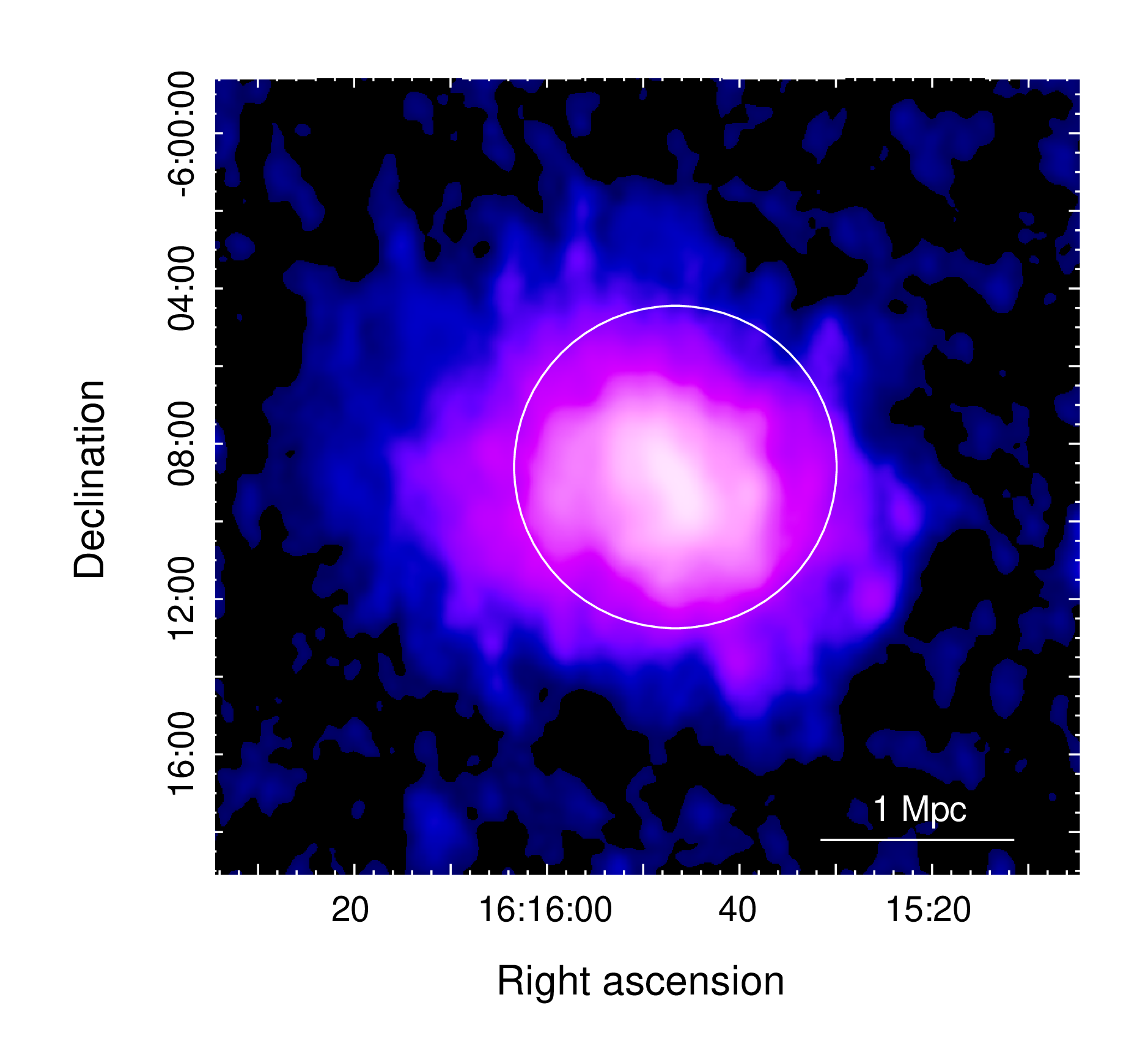}
\caption{
Radio images of A\,2163 from the combined 1.4 GHz dataset. Top: radio brightness image before source subtraction. The angular resolution is  $24^{\prime\prime}\times18^{\prime\prime}$, in p.a. $-13^{\circ}$
and rms noise level is 25 $\mu$Jy beam$^{-1}$. Crosses mark the position of discrete radio sources. The tailed radio galaxies and peripheral diffuse sources identified by \citet{2001A&A...373..106F} are also labelled.
Bottom: radio brightness image of the giant radio halo after source removal. The angular resolution is  $55^{\prime\prime}\times39^{\prime\prime}$, in p.a. $4^{\circ}$ and rms noise level is 30 $\mu$Jy beam$^{-1}$.
\label{fig:radio}
}
\end{figure}

We calibrated and reduced each dataset separately using the NRAO Astronomical Image Processing System (AIPS) package.
We followed the standard calibration scheme, with amplitude and phase calibration carried out using the primary (3C286)
and secondary (1557-000) calibration sources. The flux density scale was set using the \citet{2013ApJS..204...19P} coefficients.
We applied phase-only self-calibration to each dataset to reduce the effects of residual phase errors in the data. Final images
were made using the multi-scale CLEAN algorithm implemented in the IMAGR task. After self-calibration, we combined the
C- and D-configuration data into a single data set. A final cycle of phase-only self-calibration was applied to the combined
dataset to improve the image quality. We reached an rms sensitivity level of 25 $\mu$Jy beam$^{-1}$ in the final combined
image, with a restoring beam of $24^{\prime\prime}\times18^{\prime\prime}$ (Figure~\ref{fig:radio}).

We made images using only baselines longer than 0.5~k$\lambda$ and 1~k$\lambda$ to image respectively
only the extended and compact radio sources unrelated to the diffuse radio halo. We identified 34 sources with
a peak flux density above the 3$\sigma$ level of 0.075 mJy beam$^{1}$ within a region of $\sim$2 Mpc radius
centered on the cluster X-ray centroid. These include the 3 tailed radio sources T1, T2 and T3 and 4 extended
features (D1 to D4) with no obvious optical counterpart identified by \citet{2001A&A...373..106F}. Finally, we then subtracted
the CLEAN components associated with these sources (for a total of 161 mJy) from the $uv$ data and used the
resulting data set to obtain images of the diffuse radio emission using the multi-scale CLEAN. An image restored
with a $55^{\prime\prime}\times39^{\prime\prime}$ beam is shown in Figure~\ref{fig:radio}. The image noise is 30~$\mu$Jy beam$^{-1}$. The radio spectral index was estimated to be $\sim$ 1 \citep{2004A&A...423..111F}.

The total halo flux density, measured within the $3\sigma$ isocontour is $189\pm10$ mJy, where the errors include the
flux calibration uncertainty (3\%), the image noise and the error due to the subtraction of the individual radio sources
in the halo region computed following \citet{2013ApJ...777..141C}. %A lower flux of $155\pm2$ mJy was measured by \citet{2001A&A...373..106F}.
Within our global extraction region, we measure a flux density of 90 mJy.

%\todo{Simona will need to write up this part for you -- when you send the paper to everyone, ask her for the relevant text for this section.}

\section{Background} \label{sec:background}

    One of the unique features of \nustar~is the presence of a large open mast separating the focal plane and optics modules. This causes the observatory to be susceptible to stray light, producing a gradient across the FOV. The stray light-induced background must be separated from the instrumental background, which is generally spatially uniform, since it is produced by unknown cosmic background sources of variable intensity. We isolate these components with {\tt nuskybgd} \citep{2014ApJ...792...48W}, which characterizes the background and allows background spectra and images to be generated.%This is so that local background regions can be used for any source regions.

\subsection{Components} \label{subsec:components}

    Since we have a good empirical understanding of the individual background components, we can characterize a background model from source free regions and apply the model to take into account the strengths of the background components within a source region. There are four main sources of background, each briefly detailed in this section. For a more detailed explanation see \citet{2014ApJ...792...48W}. %Wik et al. 2014.

    The internal background of the instrument comes from a few different sources. The first is the radiation environment present in \nustar's low earth orbit, which produces a flat background. Gamma rays that Compton scatter within the detector also produce background features. The rest of this component comes from fluorescence and activation lines mostly present within the range of 22-32 keV \citep{2014ApJ...792...48W}. %(Wik et al. 2014).

    The next background component is the aperture stray light, which comes from unfocused X-rays that are able to pass through the aperture stops and strike the detectors due to \nustar's open design. Because the CXB is roughly uniform throughout the sky but partially occulted by the optics bench, a gradient is produced across the FOV. The CXB spectral model used is the {\tt HEAO1 A2}-determined model \citep{1987NASCP2464..339B} %(Boldt 1987) 
    valid in the range from 3 keV to 60 keV.

    Reflected and scattered stray light is also a source of background due to the open nature of the telescope. The backside of the aperture stops in particular are a prime source of reflection that is seen by the detectors. The three primary sources of light that gets reflected are the CXB, the Sun, and the Earth. About 10-20$\%$ of the unfocused light is reflected \citep{2014ApJ...792...48W}. Solar and terrestrial light are detected in lower energy bands, particularly around 1 keV during high periods of solar activity, accounting for roughly 40$\%$ of the events below 5 keV \citep{2014ApJ...792...48W}.

    The final component is the focused cosmic background (fCXB), which is the background contribution from unresolved foreground and background sources within the field of view. Contributions from the fCXB are generally below 15 keV and amount to $\sim$10\% of the CXB photons detected, with the majority entering undeflected through the aperture stops \citep{2014ApJ...792...48W}.

\begin{figure}
\centering
\includegraphics[scale=0.75]{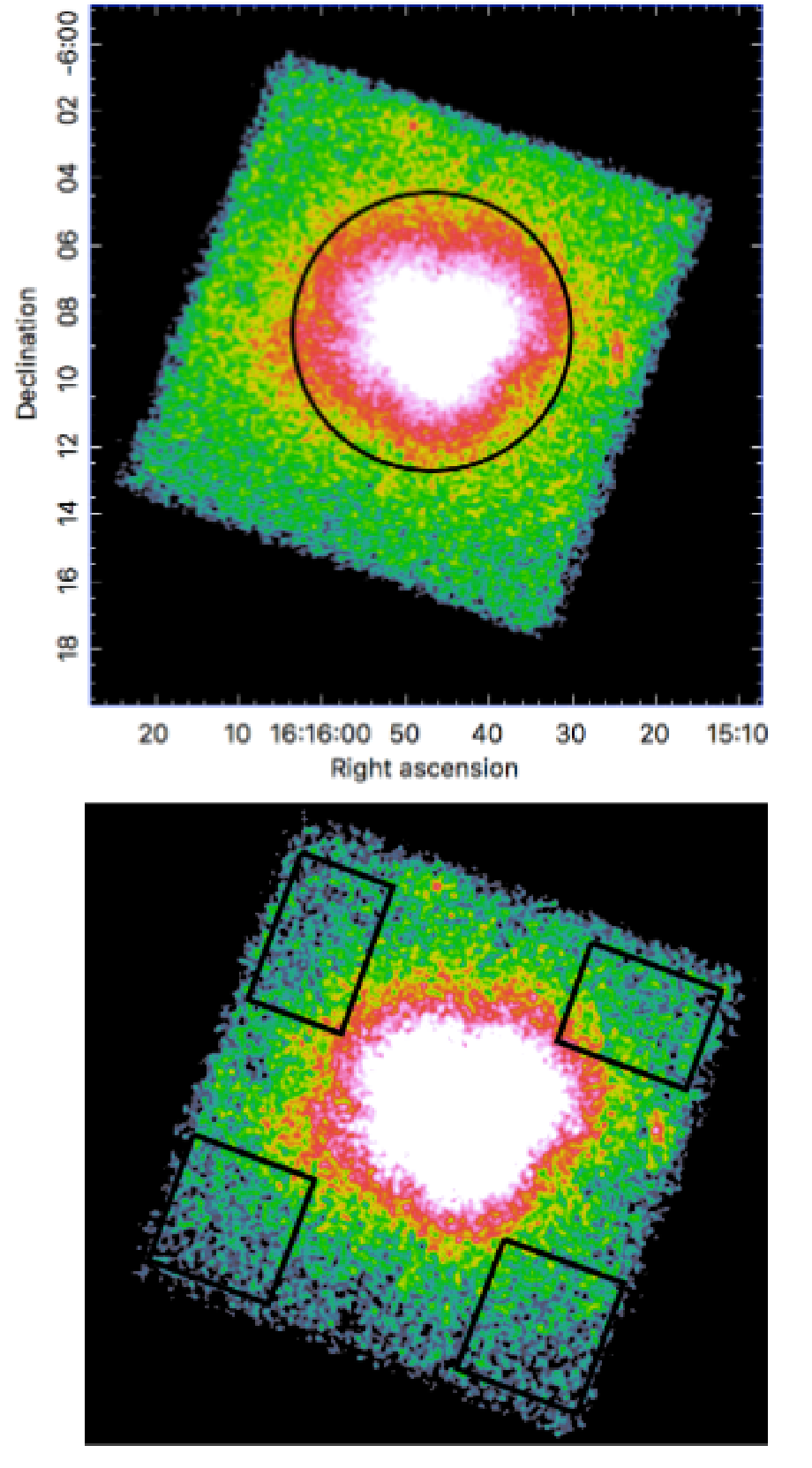}
\caption{
Top: A false color (faint to bright represented by black to blue to green to yellow to red to white) combined (A$+$B) log scaled image in the 4--25~keV band, smoothed by a Gaussian kernel with $\sigma$ = 3 pix, and stretched to show features in the outer parts of the FOV.
The source region from which spectra are extracted is shown as the black circle. A foreground galaxy group lies in the north and a background AGN is located to the west of the cluster center were kept outside the extraction region to prevent false detections of non-thermal emission. Bottom: The nuskybgd regions used to characterize the background of the cluster.}

\label{fig:region}

\end{figure}

\subsection{Background Characterization} \label{subsec:bgdfit}

\begin{figure}[h]
\includegraphics[scale=0.3785]{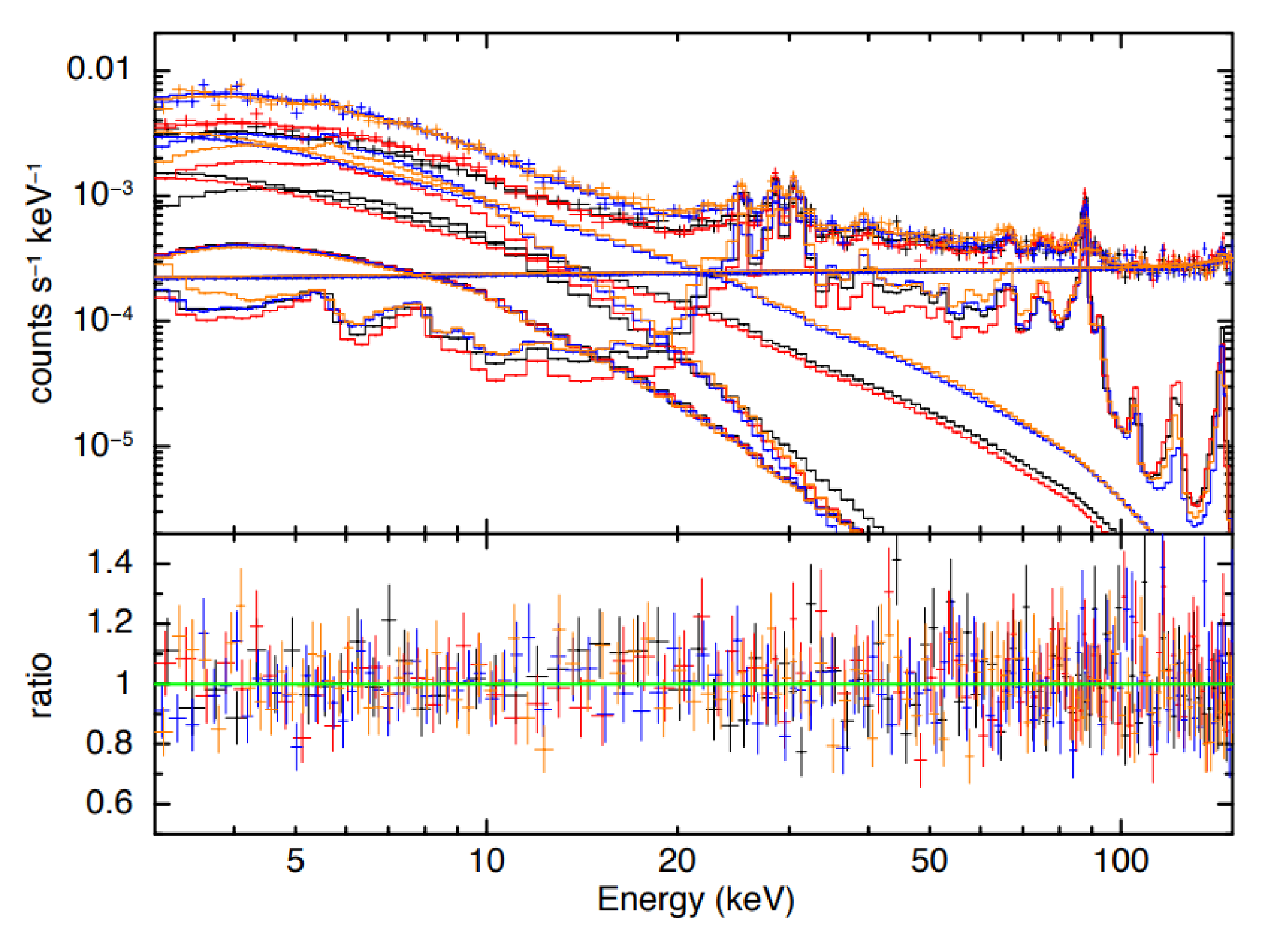}
\includegraphics[scale=0.4]{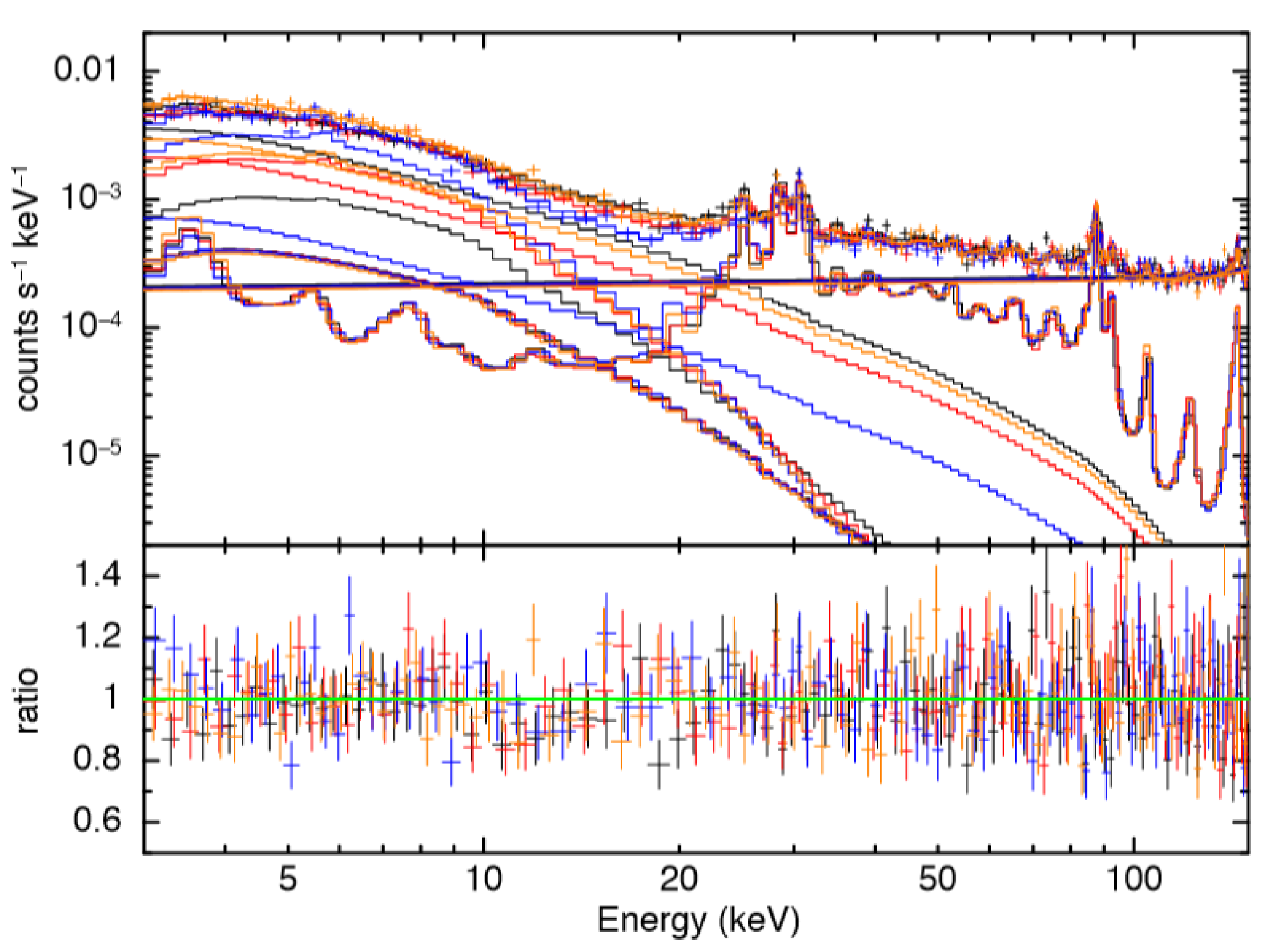}
\caption{
Fits to the background spectra extracted from the regions shown in Figure~\ref{fig:region}; the top and bottom panels show the spectra from Telescope A and B, respectively.
%A plot of the different sources of background and their respective energies. 
The background at softer energies ($<$5~keV) is primarily due to the Sun, which is combined with the model that includes instrumental background lines present at all energies. Between 10--20~keV, the undeflected CXB through the aperture stops is dominant, while the portion focused through the optics (fCXB) is $\sim$10$\times$ fainter. The steeper spectral models in this energy range are due to ICM emission from the cluster. 
%Up to 50 keV, the CXB reaches the aperture (aCXB\todo{never defined in text, either don't use or define}). 
At harder energies, cosmic ray-induced activation and fluorescence lines dominate, along with a flat continuum component representing the overall instrumental background level.
%Top: Telescope A; Bottom: Telescope B.
\label{fig:bgdspec}
}
\end{figure}

Due to the well understood nature of the aforementioned background components, we can model its composition by creating regions that are relatively source-free, as shown in the bottom panel of Figure~\ref{fig:region}. This, in combination with knowing how these components vary spatially, allows us to apply that to the source region. These source-free regions are not always completely source-free given the diffuse nature of the emission from the galaxy cluster. To take this into account, we produced a contamination file containing {\tt APEC} models used to characterize potential thermal emission due to the cluster with the temperature thawed. Results for our background fits for both Telescopes A and B can be seen in Figure~\ref{fig:bgdspec}. We find that there is no major difference observed in the background between both telescopes.

%\todo{You don't discuss how you actually characterize the background anywhere.  What regions were used?  How was contamination from cluster emission dealt with?  Figure~\ref{fig:bgdspec} isn't referenced anywhere.  Discuss that here.  You should add the regions you used to characterize the background to figure~\ref{fig:region}, or if the image would be too cluttered, add a second panel with the regions placed on the same image.}

\subsection{Systematic Uncertainties} \label{subsec:uncertainties}

    To include systematic uncertainties, we follow the procedure outlined in \citet{2014ApJ...792...48W} %Wik et al. 2014 or is it 2014?
    in their analysis of the Bullet Cluster. Due to the faint nature of an IC emission spectral component, it is paramount to accurately consider systematic uncertainties in order to prevent false detections and to derive conservative upper limits. For the instrumental background, we adopt a 90\% uncertainty of roughly 3$\%$ to account for systematic variations.
        For the aperture CXB, we adopt a systematic uncertainty of 8$\%$. 
        %\edit{This value results from the expected amount of cosmic variance given solid angle on the sky from which the stray light originates.}
        %This comes from measuring the correlation between source and non-source regions and their CXB normalizations, which are well understood. 
        The origin of this value is from cosmic variance, which is when measurements are affected by the cosmic large-scale structure (i.e., variability in flux that depends on the solid angle sampled).
        The uncertainty in the fCXB is derived in the same way, considering the smaller solid angle sampled by these sources, yielding potential variations of 42\%. 
        %This number is identical to the aCXB estimate. 

%\begin{figure*}
%\gridline{
%\fig{6.png}{0.45\textwidth}{(a) The 1T, 2T (Hot), and T$+$IC models realized 1000 times using the methods described in section 3.2.}
%\fig{15.png}{0.45\textwidth}{(b) The 2T model is significantly better at describing the temperature distribution in the galaxy cluster than the T$+$IC model.}
%}

%\gridline{
%\fig{13.png}{0.45\textwidth}{(c) The cold component of the 2T model.}
%\fig{14.png}{0.45\textwidth}{(d) The power law normalizations from the T$+$IC model shifted by the background uncertainties detailed in section 3.2 1000 times.}
%}

%\caption{
%Comparison of the different temperature models.
%}

%\end{figure*}

\section{Analysis} \label{sec:analysis}

\begin{figure}[h]
\centering
\includegraphics[scale=0.375]{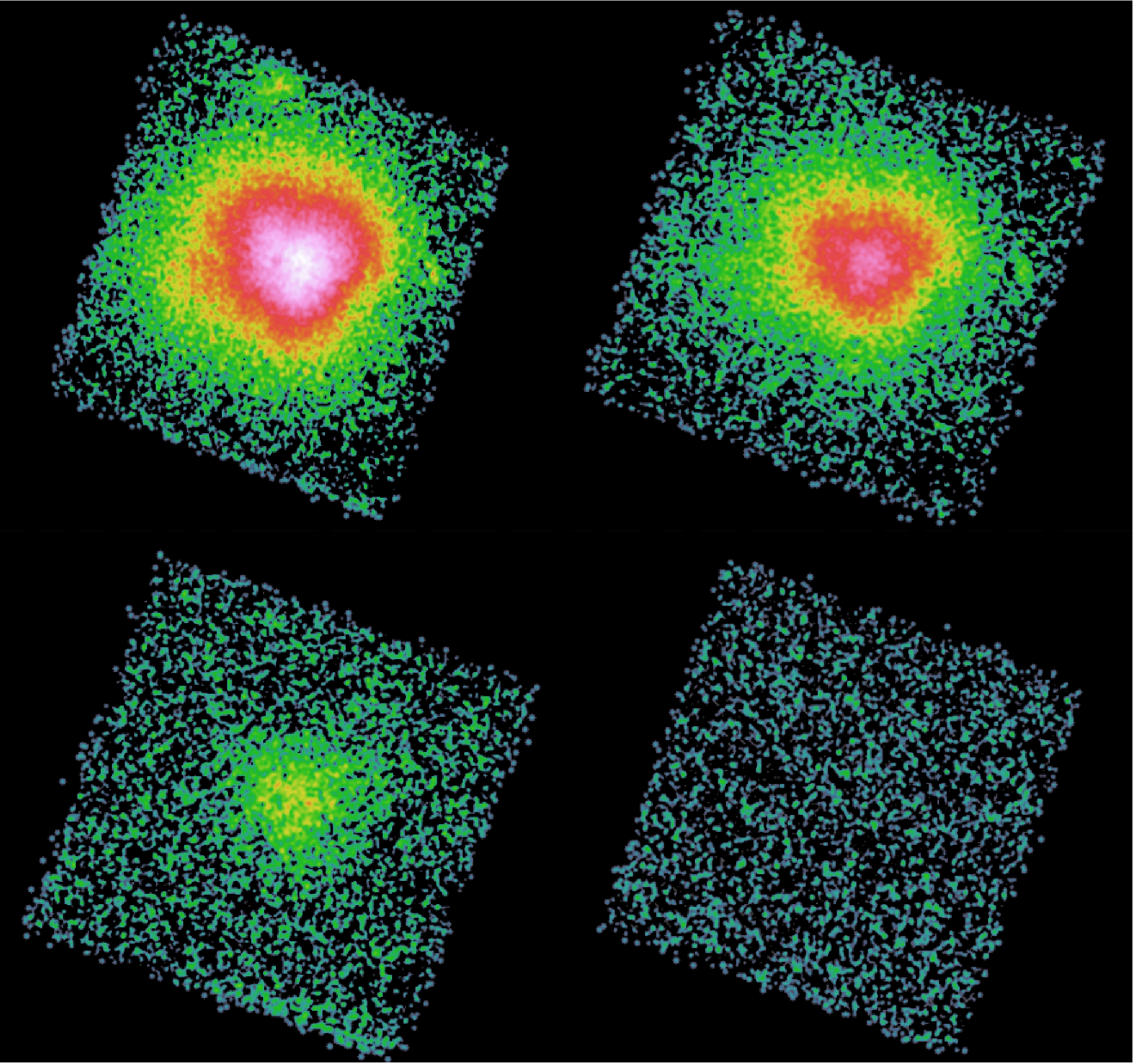}

\caption{A2163 in different energy bands. Top left: 3--8 keV; top right: 8--15 keV; bottom left: 15--30 keV; bottom right: 30--40 keV. Each image has been background subtracted and exposure corrected. They are presented in a log scale from 0 counts~s$^{-1}$~pix$^{-1}$ (in black) to 20+ counts~s$^{-1}$~pix$^{-1}$ (in white) and smoothed by a Gaussian kernel with $\sigma$ = 3~pix. %? 
There are fewer cluster counts present in the higher energy images and there are no obvious morphological changes with respect to the lower energy images, which are dominated by thermal photons.
\label{fig:energybands}
} %replace with same image but with coordinates

\end{figure}

    Before extracting spectra from the data, we must first generate images of the cluster to see if there are any bright point sources within the FOV. The post-pipeline event files can produce images in any energy band by filtering the PHA column in {\tt xselect}.
    In Figure~\ref{fig:region}, we present a smoothed image of the raw counts extracted from the combined A and B telescope cleaned event files in the 4--25~keV band. 
    We then exposure-correct and background-subtract the images with exposure and background images generated from two routines: {\tt nuexpomap} and {\tt nuskybgd}. Exposure maps are created at single energies for each band corresponding to the mean emission weighted energy of the band assuming an {\tt APEC} thermal model with $kT = $ 10 keV. Images in four different energy bands (top row: 3--8~keV and 8--15~keV; bottom row: 15--30~keV, 30--40~keV) are presented in Figure~\ref{fig:energybands}.
    %make images in these ranges
    %fix this figure number to be higher (rearrange figure order)
    From the 3--8 keV image, we see the brightest cluster emission, along with two other sources, 
    %most likely active galactic nuclei (AGN), 
    located to the north (a foreground galaxy group unrelated to A2163) and west (a background AGN) of the cluster. These sources were avoided to reduce emission that may contaminate the data and be confused with the diffuse non-thermal emission we are searching for. From the 8--15~keV image, the northern group disappears because of its soft spectrum, while the harder AGN source to the west remains. Above 15~keV, the cluster's emission begins to blend in with that of the background and above 30~keV the image becomes entirely background dominated. There are no appreciable features that may resemble non-thermal emission at these higher energies present.
    No other point sources are evident outside of the cluster, however due to \nustar's large PSF it is difficult to rule out possible sources of contamination within the brightest region of diffuse cluster emission. %More stuff in Dan's paper that I can add, not sure if this is relevant here. 
    The entire FOV of \nustar\ is a factor of two smaller than the effective PSF of {\it Suzaku}'s HXD-PIN and {\it Swift}-BAT instruments, which greatly reduces the chances of point sources (or a single bright AGN) contaminating the hard X-ray emission. Figure~\ref{fig:region} shows the source region from which we extracted the spectra that will be discussed in the following subsections. 
    
    %with the black circle being the extraction region for the spectra discussed in the next section. There are two bright point sources (most likely AGN) located to the north and west of the cluster (and indicated by the smaller black circles in Figure 8). These were excluded to avoid false positive detections of non-thermal emission. %Once you have the images, describe them individually here
    %(more stuff)

%From Figure 3 we see that there are two possible bright point sources, one to the north of the cluster and one to the west of the cluster. We create two exclusion regions around the two sources in order to remove their contributions to the spectra.

%\begin{figure}
%\includegraphics[scale=0.40]{fourtotwentyfive.png}

%\caption{\footnotesize{Our source region from which spectra were extracted.}} %replace with same image but with coordinates

%\end{figure}

\subsection{Spectra} \label{subsec:spectra}

\begin{figure}[h]
\centering
\includegraphics[scale=0.425]{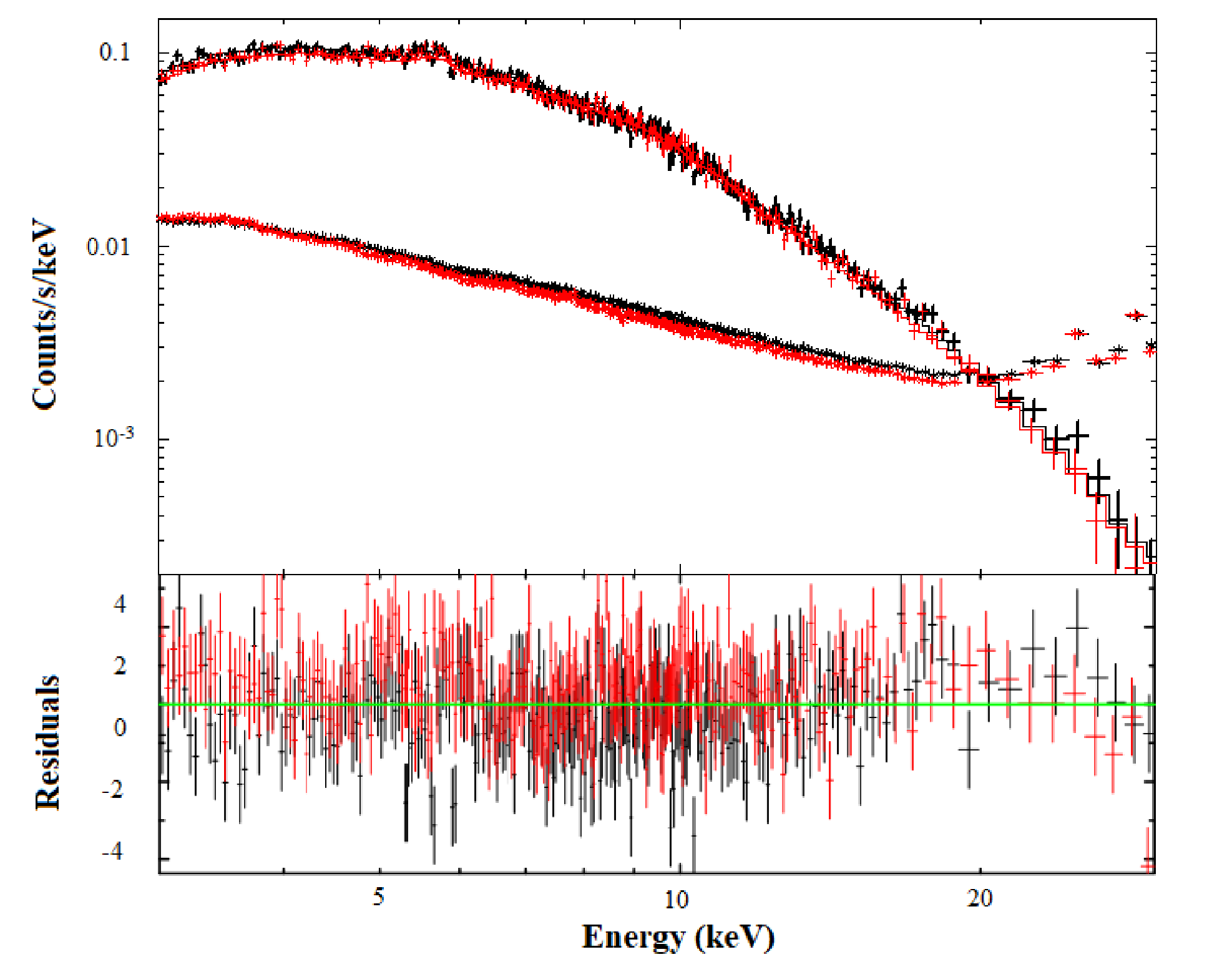} %there is an updated image of both this and the 2T model with the models shown, but the curves don't look as nice, I think it has to do with my binning or rescaling but I'll show you.
\caption{Upper panel: Background-subtracted global spectra of A2163 over the 3--30~keV bandpass fit to a single temperature (1T) {\tt APEC} model (solid lines), with the data from Telescopes A and B shown in black and red, respectively. The lower curves show the background. Lower Panel: Residuals of the fit scaled by the uncertainty in each bin.}
\label{fig:T}
\end{figure}

\begin{figure}[h]
\centering
\includegraphics[scale=0.425]{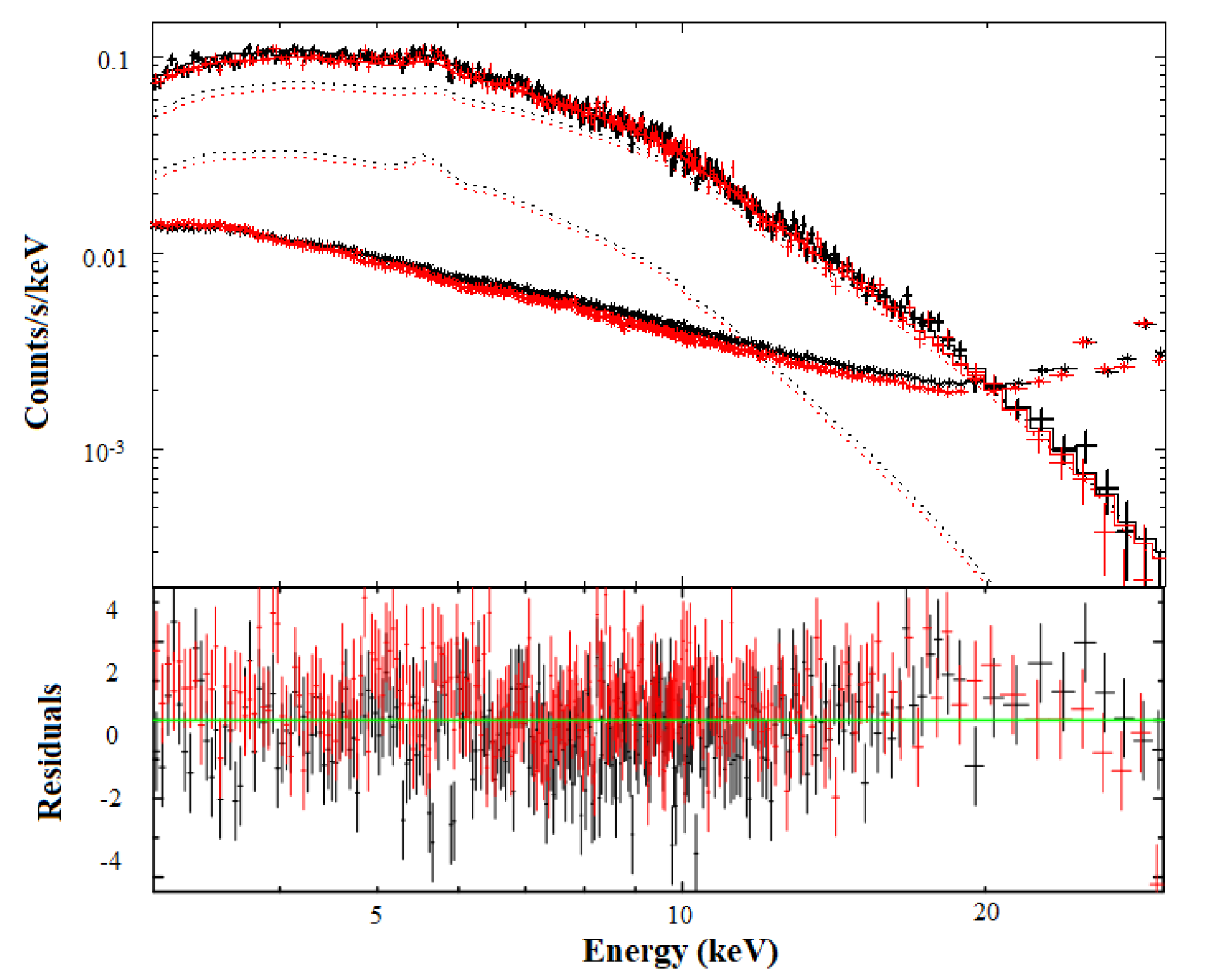}
\caption{Fit of the global spectra to the two temperature model (2T: {\tt APEC}$+${\tt APEC} model).  Note how adding the second temperature component evens out the errant residuals around the Fe complex and at high energies, compared to the 1T model.  Details are the same as in Fig.~\ref{fig:T}.}
\label{fig:TT}
\end{figure}

\begin{figure}[h]
\centering
\includegraphics[scale=0.425]{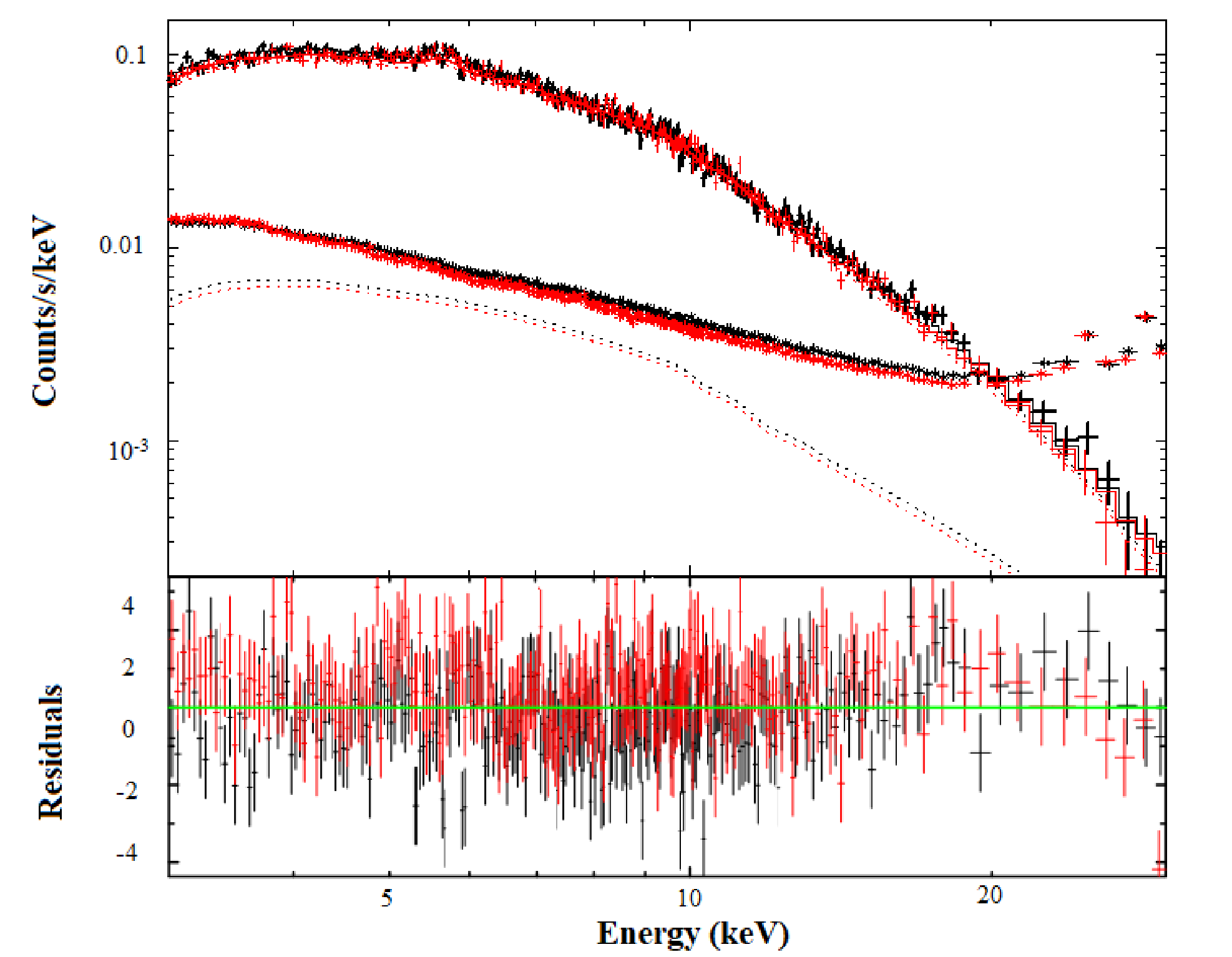}
\caption{Fit of the global spectra to the single temperature plus power law model (T$+$IC: {\tt APEC}$+$power law model).
%Background subtracted A2163 spectrum ranging from 3 to 30 keV fit to a T + IC APEC + PO model. 
The two components are shown individually as dashed lines, with the thermal component's contribution dominating over that of the IC component, which lies at or below the background level.
%The power law component, shown in the dashed lines, falls below the background.
Other details are the same as in Fig.~\ref{fig:T}.}
\label{fig:TIC}
\end{figure}

\begin{figure}[h]
\centering
\includegraphics[scale=0.425]{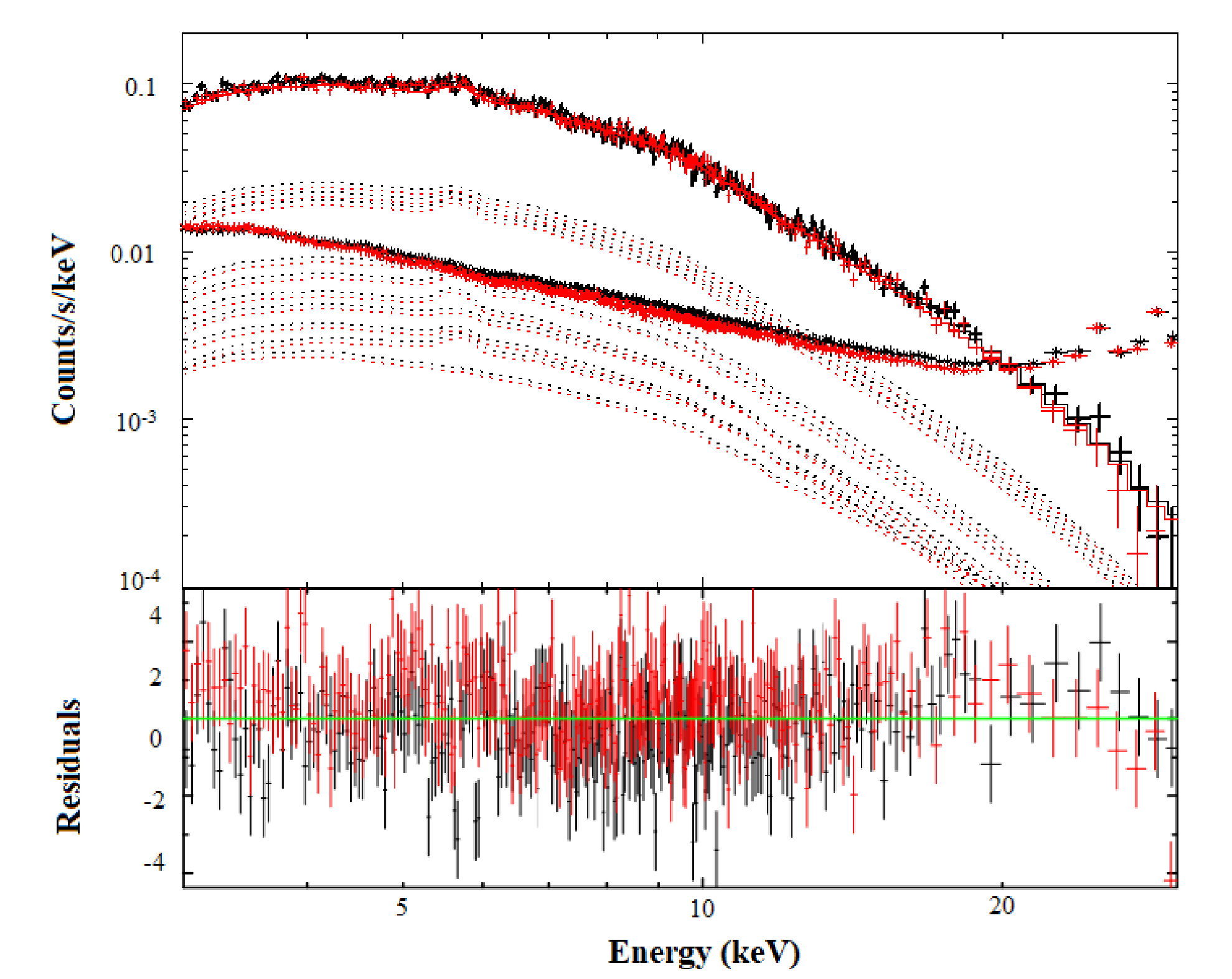}
\caption{Fit of the global spectra to the nine temperature model plus power law (9T$+$IC: 9{\tt APEC}$+$power law model, where each {\tt APEC} model is fit using the parameters in Table~\ref{table:t1}.  The lower dashed line is the non-thermal power law component, with all the dashed lines above it being brigher thermal components. Details are the same as in Fig.~\ref{fig:T}.
\label{fig:9TIC}}
\end{figure}

Spectra were generated with {\tt nuproducts} from the region shown in Figure~\ref{fig:region} and fit with {\tt XSPEC}, with the fitted spectra shown in Figures~\ref{fig:T},~\ref{fig:TT},~\ref{fig:TIC}, and~\ref{fig:9TIC}. %Should I put a raw spectrum here?
The two spectra from the A and B telescopes are generally consistent with each other. %Background emission within the source region was taken into account by an empirical model created from blank field observations fit to source excluded regions. 
To include the impact of systematic uncertainties due to the background, we simulated 1000 different realizations of the individual components of the background described in section~\ref{subsec:components} randomly shifted by an amount consistent with the variances quoted in section~\ref{subsec:uncertainties}, similar to the approach taken by %Moretti et al. (2011) and 
\citet{2014ApJ...792...48W}. %Wik et al. (2014). 
The fluctuations are assumed to follow a Gaussian distribution.
New background spectra are generated for each iteration, subtracted from the source spectra, which are grouped by 30 counts, and best-fit parameters are found using
the modified Cash statistic (via the {\tt statistic cstat} command in {\tt XSPEC}). This was done to speed up the fitting process while also not creating an appreciable loss of information due to the larger bins (since the continuum components are of greatest interest).
%Fits are made to each 1000 iterations with the {\tt statistic cstat} {\tt XSPEC} command in order to avoid bias from the lack of counts per bin that would be present if $\chi$$^2$ were applied instead. 

%\begin{figure}
%\includegraphics[scale=0.4]{okbackground.PNG}
%\caption{\footnotesize{A plot of the different sources of background and their respective energies. Softer energies (< 5 keV) primarily come from the Sun. Between 10 keV and 20 keV, the CXB focused through the optics (fCXB). Up to 50 keV, the CXB reaches the aperture (aCXB). At energies harder than that, cosmic ray radiation and fluorescence produce internal lines. Left: Telescope A; Right: Telescope B.}}
%\end{figure}

\subsection{Models} \label{subsec:models}

    In clusters with radio relics and radio halos like A2163, there must exist IC emission at some level. In the case where it is weak, such as with this cluster, the total model used to fit data becomes more important in differentiating non-thermal from thermal emission. Following the methodology done in \citet{2014A&A...562A..60O} %Ota et al. 2014 
    for A2163 and \citet{2014ApJ...792...48W} %Wik et al. 2014 
    for the Bullet Cluster, we use the four following models: single temperature (1T), two temperature (2T), single temperature added to a power law (T$+$IC), and multi-temperature added to a power law (9T$+$IC). The temperature components are all calculated using the {\tt APEC} model within {\tt XSPEC}. It should be noted that metal abundances are allowed to be free during the fitting process and that we ignore foreground absorption, due to \nustar's lack of sensitivity below 3~keV. 
    %has negligible affect on our results. 
    We verified that the latter has negligible affect on our results by multiplying all of our fits by the {\tt tbgas} model, which takes as an input the Galactic absorption. Using values for $n_H$ of 0.118 \citep{2005A&A...440..775K} and 0.230 (based on results from \citet{2013MNRAS.431..394W}, we found no statistically significant differences between the fits containing {\tt tbgas} and those lacking it. All model parameters 
    %are summarized in Table~\ref{table:T2} and 
    are discussed in more detail in the following subsections.

\subsubsection{Single Temperature} \label{subsubsec:T}
    The 1T model takes a very simplistic approach to characterizing the total emission from the cluster. If the gas is nearly isothermal and entirely thermal in nature, this model could provide a satisfactory fit given that \nustar's 0.4 keV FWHM resolution limits our ability to separate the 6.7 keV He-like and 6.9 keV H-like iron lines---the only significant emission lines in \nustar's bandpass at these temperatures. 
    % This belongs in the 2T section, not here
    %Data taken by \xmm, however, suggests that the gas within the ICM varies in temperature \citep[e.g.,][]{2004ApJ...605..695G}, \edit{consistent with merger activity}. 
    %\edit{Despite previous such measurements}, the true temperature structure of the cluster is unknown. %but for the global spectrum this is not a problem. 
    %Instead, these measurements provide emission-weighted temperatures potentially biased by a telescope's effective area and gas projections along the line of sight. This means that the temperature is dependant on the energy range and calibration of the telescope. To measure the temperature distribution detected by \nustar, a 2T model provides a way to take into account that \nustar\ is preferentially weighted towards hotter temperatures and may not agree with projected temperature structure measured at lower energies. Due to the lack of features in the thermal continua, the suggested 2T model can account for a much larger range of temperatures. 
    Previous work using {\it XMM-Newton}$+${\it Suzaku}~HXD-PIN data gave a best 1T model fit of $14^{+6}_{-5}$~keV \citep{2014A&A...562A..60O}. %(Ota et al. 2014). 
    In contrast, our 1T fit to \nustar\ spectra results in $kT = 11.8\pm0.2$~keV (statistical uncertainty only), consistent with the previous estimate but much more precise. The C-stat value for this fit is 1434 with 1344 degrees of freedom (dof).
    %\todo{Weird abbreviation choice -- I think ``dof" is common, but be consistent everywhere.  In the table you use ``dof.", which makes more sense than leaving off the last dot here (which you do elsewhere in the text, so it's not a typo).}

\subsubsection{Two Temperature} \label{subsubsec:TT}
    Spatially-resolved temperature estimates from \xmm, however, suggest that the gas within the ICM of A2163 is not isothermal \citep{2004ApJ...605..695G, 2011AA...527A..21B}, consistent with ongoing merger activity. 
    Despite previous such measurements,
    the true temperature structure of the cluster is unknown. %but for the global spectrum this is not a problem. 
    Instead, these measurements provide emission-weighted temperatures potentially biased by a telescope's effective area and gas projections along the line of sight. This means that the temperature is dependant on the energy range and calibration of the telescope. To better measure the temperature distribution detected by \nustar, a 2T model provides a way to take into account that \nustar\ is preferentially weighted towards hotter temperatures and may not agree with projected temperature structure measured at lower energies. Due to the lack of features in the thermal continua, a 2T model can in principle account for a range of temperatures.
    
    %As mentioned before, the 2T model allows for a more accurate representation of the harder X-ray band spectrum.
    %what do you mean pick???
    %We pick a 2T model based on the bimodal distribution of temperatures in the Bullet Cluster \citep{2007ApJ...670.1010A, 2014ApJ...792...48W}. %(Andersson et al. 2007, Wik et al. 2014). 
    We find the values for the two temperature components to be $T_{\rm h} = 13.5^{+2.2}_{-1.1}$
    %\pm 6.5, 3.4 check these errors again  %A fit to this model %finish
    and $T_{\rm l} = 5.9^{+3.8}_{-3.1}$~keV. The C-stat value for this model is 1411 with 1342 dof, suggesting a better fit than the 1T model.
    If significant IC emission were detectable, it should be noticeable in this model through the hot temperature component, which would be unphysically high. If this is the case, then using a T+IC model would provide a better fit to the overall spectrum. As is the case with the 1T model, the thermal component would not entirely follow the actual thermal distribution, but the shape of the spectrum at higher energies would be better described by the non-thermal power law component. 
    %\pm 3.2, 2.9 check these errors again
    %We also attempted to fit a nine temperature model based on the regions in figure 7. The normalization for this model was very close to one. %finish later
    %Include 9T here or make a new section?
    
\subsubsection{T$+$IC} \label{subsubsec:TIC}   
    The T$+$IC model is a more accurate representation of the spectrum in the case when the non-thermal component is particularly prominent. In our limited bandpass, IC scattering follows a power law curve, so we use {\tt XSPEC}'s power law model with a photon index of $\Gamma = 2$, since our $\alpha$ value based on the radio data was $\sim$ 1, roughly following the best fit value found by \citet{2014A&A...562A..60O} of 2.18. We attempted to fit with a free photon index, but when done it would become unphysically steep. This is likely due to the photon index mimicking the effects of $T_l$ in the 2T model.  It is important to note that due to declining counts at harder energies, a non-thermal component due to IC scattering needs to be strong enough to surmount worsening the fit at softer energies, where the majority of counts are located. %See Wik et. al 2014 section 4.2.1 last sentence in paragraph 2.
    Here, our thermal component and the power law normalization were left free. The best-fit temperature obtained from the model is $kT = 11.4\pm0.3$~keV, lower than just a single temperature model, which makes sense due to the harder non-thermal component accounting for some of the photons at the highest energies. The estimated power law flux from 20--80~keV for the T$+$IC model is $F_{\rm NT}$ $<$ 4.03 $\times$ $10^{-12}$~\flux~with a 90$\%$ confidence level. 
    %This is used because the spectrum gets folded by the ARF, so the photons may not register at their original energies but as a spread. 
    %{\tt XSPEC} does a matrix equation where it uses the number of events (RMF $\times$ ARF) and a model to tell you how many photons you are getting out. This allows you to get the flux from the actual incident photons, which would be close, but not quite exact due to the bin size, which may prevent the integration from working properly. Using {\tt dummyrsp} makes a diagonal response matrix that makes smaller bins and makes it so that the number of counts is roughly the number of photons. 
    %Using this value, we obtain a magnetic field strength lower limit of B $>$ 0.22 $\mu$G. Repeating the fit using our 9T regions results in an upper limit for the non-thermal flux of $F_{NT}$ $<$ $1.62 $\times$ 10^{-12}$~erg~$s^{-1}$~$cm^{-2}$ and a lower limit on the strength of the magnetic field of B $>$ 0.35 $\mu$G %(MORE SENTENCES!). Figure out errors for fluxes, recalculate fluxes and magnetic fields for 1T, 2T, and 9T Should I talk about magnetic field strength here as well?
    We do not attempt a 2T+IC model, which in theory should better constrain the thermal components as mentioned before. This would provide a better constraint on a non-thermal excess at higher energies while minimizing the effects on the errors of the thermal parameters. However when doing this with A2163, one of the two thermal components gives a very low, unphysical temperature. %given it's extrapolation to energies below 3 keV. 
    Since this component is simply making small corrections to the model at the lowest energies, where uncertainties in the background or overall calibration are larger, the addition of a second temperature component
    %While this serves as a useful correction for the entire thermal spectrum which is largely dominated by the hotter thermal component, it is poorly constrained and thus 
    provides no advantage over the T+IC model, which adequately accounts for the average thermal emission given the quality of the data. The C-stat value for this model is 1419 with 1343 dof, which means that this model does not fit the data as well as the 2T model. This signifies that the multi-temperature nature of the spectrum dominates over the need for a non-thermal component driven by an excess at hard energies.

   % \todo{It is significantly worse in fact -- might be worth commenting more about what this means (that the multi-temperature nature of the spectrum dominates over the need for an IC component driven by an excess at hard energies.}

\subsubsection{Multi-temperature + IC} \label{subsubsec:9T}
\begin{figure}[h]
\centering
\includegraphics[scale=0.375]{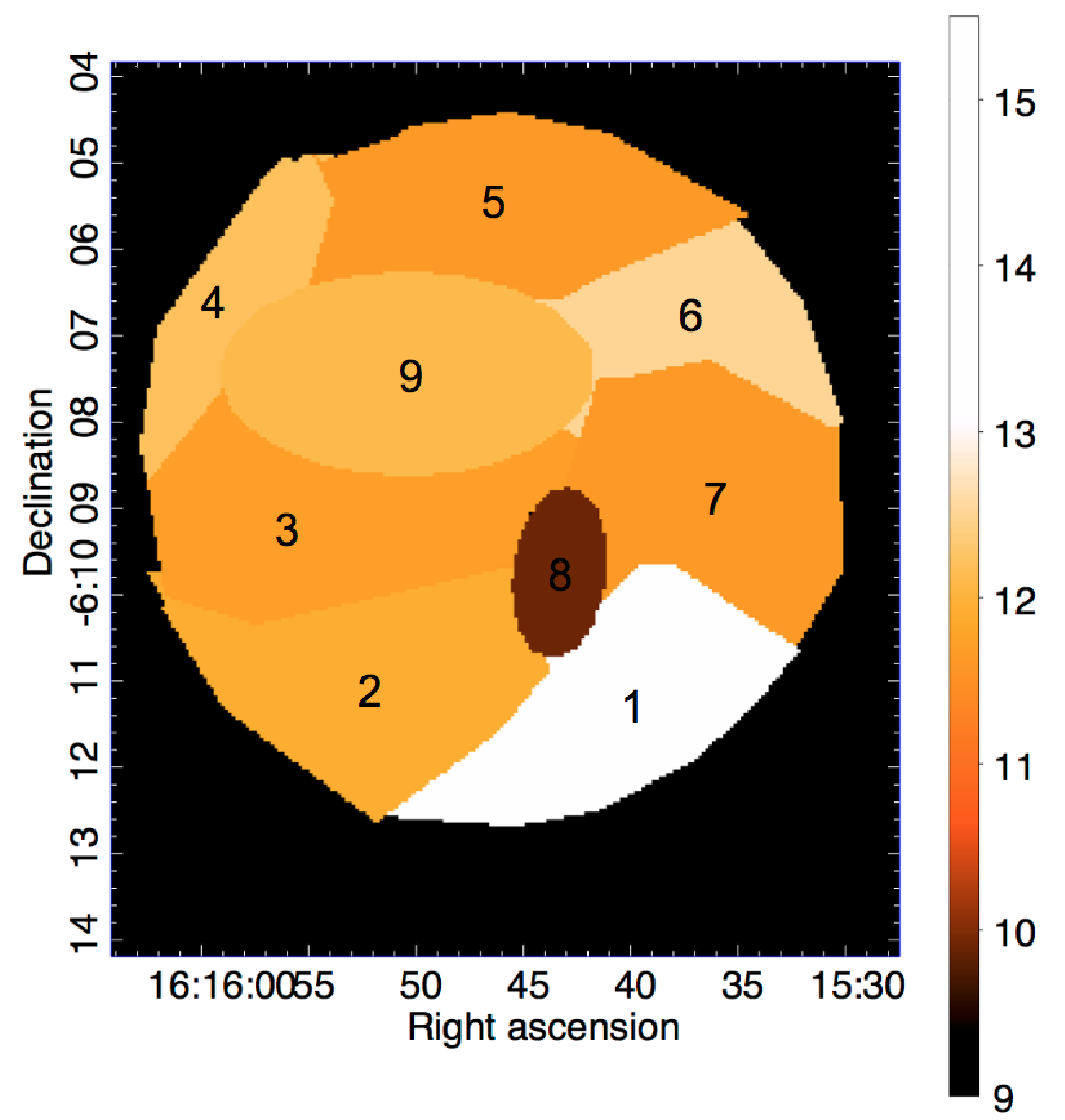}
\caption{
A temperature map inside the extraction region of our global spectra, consisting of 9 regions selected based on the work of \citet{2011AA...527A..21B}. We do not detect higher temperature gas in the northeast (in region 9) as seen in both \citet{2011AA...527A..21B} and \citet{2014A&A...562A..60O}.
The discrepancy is likely due to spatially variable absorption, which \nustar\ is insensitive to.
Temperature variations are primarily due to the disrupted cool core (region 8) and a region of hot gas likely heated by a shock front driven ahead of the cool core to the southwest (region 1).
See Table~\ref{table:t1} for temperatures, abundances, and normalizations.}
\label{fig:temperaturemap}
\end{figure}

Given the spatially-resolved spread in temperatures in A2163, a model containing multiple temperature components and a non-thermal power law component should more accurately capture the emission at all X-ray energies. 
    %It is important to note, however, that using 
    As found in the previous section, however, any model containing more than two free temperature components will become degenerate or unphysical. In this situation, a physically meaningful better fit will not be obtained due to the
    %This is partially due to temperatures being skewed by photons from neighboring regions and the fact that there are more parameters and 
    fewer degrees of freedom. 
    To incorporate our knowledge of the true multi-temperature nature of A2163's ICM, and to take advantage of \nustar's spatial resolution, we extract spectra from 9 smaller regions within the global extraction region and fit them individually with single temperature models.
    We base these regions on those presented in the temperature map of \citet{2011AA...527A..21B}, which are shown in Figure~\ref{fig:temperaturemap}, and the best-fit model parameters for each region are given in Table~\ref{table:t1}. When compared to the \xmm~temperature map, we do not observe the 18~keV, shock-like feature in the northeastern corner of the cluster (corresponding to our region~9).
    %\todo{In the below instance and in general, it is grammatically incorrect to not follow ``this" with the noun being referred to.  For example, ``This means that" is a terrible way to start a (written) sentence.  Instead, indicate what ``this" is referring to: ``this poor fit means that...", ``this result implies..."  I fixed a bunch of these in my earlier pass.}
    This difference is likely due to A2163's placement with respect to the Galactic plane. The cluster lies close to the plane containing high amounts of neutral hydrogen. If the distribution of the column density varies on small angular scales, but is taken to be constant as done in the \xmm~study, the temperatures can be skewed to higher or lower values by the fit as the fit statistic at lower energies is more highly weighted than in the higher energy portions of the spectra. \nustar\ does not run into this issue due to its insensitivity to foreground absorption.  
    %It is for this reason that we use regions fitted individually with single temperatures and then sum them up.
    
    The resulting best-fit parameters are then used to construct a composite 9T thermal model that in principle better represents the thermal structure contained within the global spectrum.
    We combine this model with an IC component (9T$+$IC) twice, once keeping all the parameters that were found individually fixed other than the overall normalization and once allowing the temperatures to vary within their error range.  
    %There was very minimal and inconsequential 
    The difference between these two methods was not significant, so we only report the former process.
    %, which is slightly more conservative.  
    %we model the spectrum twice, once with fixed parameters based on the individual fits of the regions and once while allowing the temperatures to be free within their error bars.  %This is so that  The average temperature of the nine regions comes out to (put average temperature which is very similar to the 1T model's global temperature.)
    With the temperatures able to vary within their uncertainties, any weak non-thermal emission present that may bias the temperature estimate high in the single temperature fit is allowed to adjust in the fit to the global spectrum, where the IC component should be more significant.
    The normalization constant for this model was very close to one, meaning that the fit parameters were comparable to the parameters obtained when fitting the regions individually. Repeating the process for determining the non-thermal flux in the previous section, we obtain a flux of $F_{\rm NT}$ $<$ 1.64 $\times$ $10^{-12}$~\flux. The C-stat value of this model is 1418 with 1345 dof, again suggesting that the best-fit model for the data is the 2T model.\\
    \indent It is possible that this method could have underestimated the presence of an IC component; i.e. the temperatures may be overestimated if hard IC emission is present, masquerading as hotter thermal flux in the model. In order to check for this possible bias, we ran a simulation test where we artificially inserted twice the amount of IC flux found in our upper limit, to check whether that level of IC emission would be missed by our approach or not. First, for each of our 9 regions, we create a 1T+IC model; the parameters of the thermal component are the same as those we found in our fits to the real data. The IC flux is added equally to each region, each with a photon index of 2 and a flux equal to two-ninths of our 90\% upper limit. Spectra are simulated and then fit with a single temperature model. Following the procedure for the real data, we use the best-fit parameters from these single temperature fits to create a 9T model. We then add the 9 simulated spectra to create a global spectrum, which we fit with the 9T model plus a power law component in exactly the same way that we fit the real data with the 9T+IC model. Were IC emission being suppressed because it is being taken up by the thermal models, we should measure a lower IC flux from the global simulated spectrum than we included. However if the thermal models are not significantly biased by this level of underlying IC emission, then we should accurately measure the total amount of IC flux we added to the simulated spectra.\\
    \indent The temperatures measured of each simulated spectra were $\sim$0.2 keV higher than the true values with the exception of the cool core, which was about 2 keV higher. The global fit found an IC flux consistent with the injected flux detected at $3.9\sigma$ significance. Therefore we conclude that an IC flux at this level would have been detected with the detailed thermal modeling approach, and that the approach would not have led to a biased estimate of the flux. Since we estimate the presence of IC emission to be weaker and thus the temperatures estimated in individual regions to be less biased, the 9T+IC model provides an accurate and unbiased way to estimate the limit on the level of IC emission in A2163.
%\todo{GAH!  I just realized you don't have a figure for the spectral fit of the 9T+IC model.  You really should include this fit as well, in which case it may make sense to put all the spectra into a single 4 panel figure -- but this can wait until after sending the draft to coauthors.}    
    
\begin{deluxetable}{cccc}
\tablewidth{0pt}
\tablecaption{This table contains the individual temperatures and abundances measured within the 9 selected regions from our multi-temperature model. These regions were chosen loosely based on regions done in previous work to create a temperature map done by \citet{2011AA...527A..21B}. %\todo{fix Bourdin citation} %Bourdin et al. 2011. 
\label{table:t1}
}
\tablehead{
& \colhead{Temperature} & \colhead{Abundance} & \colhead{Norm\tablenotemark{1}} \\
\colhead{Region} & \colhead{(keV)} & \colhead{(Solar)} & \colhead{($10^{-3}$)}
}
\startdata
1 & $15.55 \pm 0.83$ & $0.51 \pm 0.21$ & $1.41 \pm 0.04$ \\
2 & $11.90 \pm 0.32$ & $0.37 \pm 0.10$ & $3.01 \pm 0.05$ \\
3 & $11.67 \pm 0.20$ & $0.31 \pm 0.06$ & $7.74 \pm 0.05$ \\
4 & $12.21 \pm 0.78$ & $0.22 \pm 0.15$ & $9.54 \pm 0.04$ \\
5 & $11.67 \pm 0.64$ & $0.22 \pm 0.14$ & $1.53 \pm 0.03$ \\
6 & $12.54 \pm 0.51$ & $0.19 \pm 0.11$ & $1.67 \pm 0.07$ \\
7 & $11.65 \pm 0.30$ & $0.21 \pm 0.08$ & $3.23 \pm 0.05$ \\
8 & $\phn9.95 \pm 0.22$ & $0.35 \pm 0.06$ & $2.44 \pm 0.04$ \\
9 & $12.11 \pm 0.20$ & $0.30 \pm 0.04$ & $8.63 \pm 0.07$ 
\enddata
\tablenotetext{1}{Normalization of the APEC model, given by $(10^{-14}/[4 \pi (1+z)^{2} D^{2}_{A}]) \int n_{e} n_{H} dV$ where $z$ is the redshift, $D_{A}$ is the angular diameter distance, $n_{e}$ is the electron density, $n_{H}$ is the ionized hydrogen density, and $V$ is the volume of the cluster.}
%\label{table:t1}
\end{deluxetable}


%We fix the power law photon index to 2, in between the best fit values found by Ajello et al. (2010) and Ota et al. (2012), which were 1.86 and 2.18 respectively.
   
\subsubsection{Preferred Model Including Systematic Uncertainties} \label{subsubsec:best}
%Distribution of temperatures better captured in 9T, 2 free params vs 5 free params in 2T however, but C-stat difference is greater than 3. Global region arf is average for entire region, 9T arfs specific to individual regions, so not entirely capturing differences precisely. The shape

\begin{deluxetable*}{cccccccc}
\tabletypesize{\scriptsize}
\tablewidth{0pt}
\tablecaption{This table contains the results of our fits using the 1T, 2T, T$+$IC, and 9T$+$IC models. The redshift for all fits was allowed to be free (nominally 0.203) to a value of $z = 0.209$. See Table~\ref{table:t1} for individual temperature components in the 9T model. Errors are presented as statistical followed by systematic.
%\todo{Several issues to fix: (1) change the 2T abundance to an upper limit, (2) the 2T normalizations are wrong.  The $T_l$ norm should be an upper limit if that error bar is correct, and the 2nd norm is NOT the norm of the hot component (or vice versa), (3) is the thermal norm for the 9T$+$IC model the summed APEC norm for all the regions?  if so, you should say this in the text -- also, this error bar seems much too large, shouldn't it be comparable to the norm on the T$+$IC component?  I would believe an error of 0.03 if the parameter value was $\sim$1 -- are you actually reporting the constant and not the norm?, (4) I am skeptical these error bars are all symmetrical -- if the upper/lower ranges do not equal, then you need to report them as $val^{upper}_{lower}$ here.}
%\label{table:1}}
\label{table:T2}
}
\tablehead{
& \colhead{Temperature} & \colhead{Abundance} & \colhead{Norm\tablenotemark{a}} & \colhead{$kT$ or $\Gamma$} & \colhead{Norm or IC Flux\tablenotemark{b}} & & \\
\colhead{Model} & \colhead{(keV)} & \colhead{(Solar)} & \colhead{($10^{-2}$~cm$^{-5}$)} & \colhead{(keV or ...)} & \colhead{($10^{-2}$ or $10^{-12}$~erg~s$^{-1}$~cm$^{-2}$)} & \colhead{C-stat\tablenotemark{d}} &
\colhead{dof}
}
%\tablefootnote{Normalization of the MeKaL thermal spectrum, given by {$10^{-14}$/[4\pi$(1+z)^{2}$$D^{2}_{A}$]}}\int$n_{e}$$n_{H}$dV where z is the redshift, $D_{A}$ is the angular diameter distance (cm), $n_{e}$ is the electron density (cm$^{-3}$), $n_{H}$ is the ionized hydrogen density (cm$^{-3}$), and V is the volume of the cluster.}
%\tablefootnote{20 to 80 keV}
%\tablefootnote{photons/keV/cm$^{2}$/s at 1 keV} do I need this if I'm just reporting IC Flux instead of the power law norm?
\startdata
1T & $11.8 \pm 0.2, 0.2$  & $0.31 \pm 0.03, 0.02$ & $3.0 \pm 0.2, 0.1$ & ... & ... & $1434^{+120}_{-116}$ & 1344 \\
2T & $\phn5.9^{+3.8, +2.1}_{-3.1, -1.9}$  & $0.48^{+0.63, +0.12}_{-0.21, -0.09}$ & $6.7^{+5.4, +1.1}_{-3.5, -1.3}$ & $13.5^{+2.2, +2.2}_{-1.1, -2.7}$ & $2.4^{+0.2, +0.2}_{-0.1, -0.1}$ & $1411^{+117}_{-121}$ & 1342 \\
T$+$IC & $11.4 \pm 0.3, 0.2$ & $0.38 \pm 0.05, 0.02$ & $2.9 \pm 0.1, 0.2$ & 2 (fixed) & $4.03^{+0.72, +0.88}_{-0.61, -0.45}$ & $1419^{+121}_{-118}$ & 1343 \\ 
9T$+$IC & ... & ... & $1^{+0.04, +0.01}_{-0.03, -0.01}$\tablenotemark{c} & 2 (fixed) & $1.64^{+0.81, +0.52}_{-0.93, -0.61}$ & $1418^{+122}_{-119}$ & 1345
\enddata
\tablenotetext{a}{Normalization of the {\tt APEC} model, defined the same as in Table~\ref{table:t1}.}
\tablenotetext{b}{20--80~keV}
\tablenotetext{c}{Normalization constant for the nine models.}
\tablenotetext{d}{Distribution of C-stat values from the 1000 realizations shown.}
\end{deluxetable*}

%13.5^{+2.2}_{-1.1}$
    %\pm 6.5, 3.4 check these errors again  %A fit to this model %finish
    %and $T_{\rm l} = 5.9^{+3.8}_{-4.1}

%old 9T abundance $0.20 \pm 0.17

\begin{figure}
\centering
\includegraphics[scale=0.80]{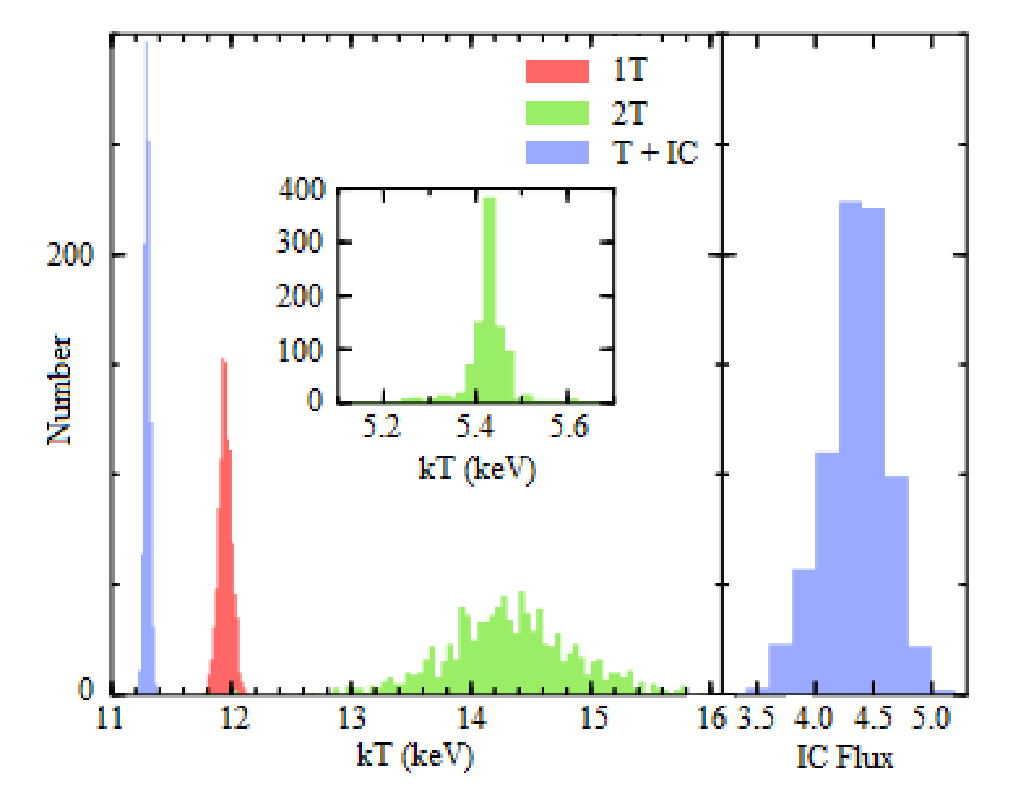}

\caption{The distribution of best-fit parameter values for the 1T (red), 2T (green), and T$+$IC (blue) models using the 1000 realizations of the background (as described in Section~\ref{subsec:uncertainties}). Parameters shown are the temperatures of each model and the IC flux (from 20--80 keV) in units of $10^{-12}$~\flux. The width of the distributions shows the effects that systematic uncertainties in our modelling have on these parameters, which are also reported in Table~\ref{table:T2}.}
%\todo{place units in (): $kT$ (keV), and report IC 20--80 keV flux instead of norm: IC flux ($10^{-12}$~\flux (20--80~keV) -- if this all doesn't fit, describe more fully in the caption}} %replace with same image but with coordinates
\label{fig:histograms}

\end{figure}

To briefly summarize, in the previous section we discussed how we arrived to our parameter values and statistical errors (shown in Table~\ref{table:T2}). 
%\todo{move the table closer to this section so the reader doesn't have to scroll so far to refer to it}
What we found was that based on C-stat values, the 2T model (1411 with 1342 dof), was the best-fit model for the data. From a physical standpoint, this makes sense when compared to the 1T model (C-stat value of 1434 with 1344 dof), as the temperature structure within a merging cluster should be more complex than that of a 1T model. The T+IC model, however, cannot be completely ruled out just based on C-stat alone. In this section, we will provide further methods for ruling out T+IC as potentially being the best-fit model. As mentioned in section~\ref{subsec:uncertainties}, we also have to include how background systematics may affect our results. Although the various components of the background (instrumental, aCXB, and fCXB) were characterized, that characterization may not have been perfect, and there is a known level of systematic uncertainty associated with each component as described in Section \ref{subsec:uncertainties}. To include these uncertainties in our analysis, we create 1000 realizations of the background, where each realization consists of the normalizations of these 3 components randomly shifted from their nominal values; the random shifts follow a normal distribution with the width of their systematic uncertainty. %To do this, we created a distribution of best-fit temperatures for the first three models discussed by randomly generating 1000 realizations of the background. 
The distribution of resulting best-fit parameters is shown in Figure~\ref{fig:histograms}. Here we can see how our choice of model and the background uncertainties affects how we can assess our spectrum. In the 1T case, where the shape of the model solely depends on one parameter, background uncertainties have a minimal effect on the temperature. This fact is reflected in the distribution of temperatures (the red histogram in Figure~\ref{fig:histograms}) being about 0.2 keV, comparable to our statistical error shown in Table~\ref{table:T2}. 
    The introduction of another parameter that can control the shape of the model, as is the case in the 2T or T+IC models, causes the background to play a larger role. In the case of the 2T model, the two best-fit temperatures are more sensitive to changes in the background due to the fact that the model has greater capability to adjust to small changes in the shape of the spectrum. The background variations mostly affect $T_h$ (as seen in the $\sim$2 keV distribution in the $T_h$ green histogram in Figure~\ref{fig:histograms}. This is because a lower or higher background will cause the spectrum to turn over at a higher or lower energy, which will in turn cause $T_h$ to go higher or lower. The $T_l$ component will then adjust to fix the lower energy part of the spectrum. This correlation means that the higher the $T_h$ temperature the higher the $T_l$ temperature. In the T+IC case (depicted in the blue histogram in Figure~\ref{fig:histograms}), the temperature, much like in the 1T case, continues to dominate the shape of the spectrum. 
    
    This is why the the temperature in this case has a similar ($\sim$0.2 keV) distribution to pure 1T case. The right panel in Figure~\ref{fig:histograms} shows the IC flux, the parameter most likely to be affected by background variations. This is because its shape is more closely resembling that of the background. 
    
    This means any shift in the background should be reflected in a shift in the normalization of the power law component in the T$+$IC model. 
    Our histogram shows that the background systematics gives an uncertainty of +0.88 $\times$ $10^{-12}$~\flux~(20--80 keV) and -0.45 $\times$ $10^{-12}$~\flux.
    If there really was a large presence of IC scattering, the background should produce a larger effect on higher energies. Instead, what we are seeing is likely similar to what the $T_l$ does in the 2T model. The IC flux is fixing the shape of the model in the lower energy range. Instead of behaving like a non-thermal component, it is behaving like a low temperature thermal component. 

\begin{figure}
\centering
\includegraphics[scale=0.75]{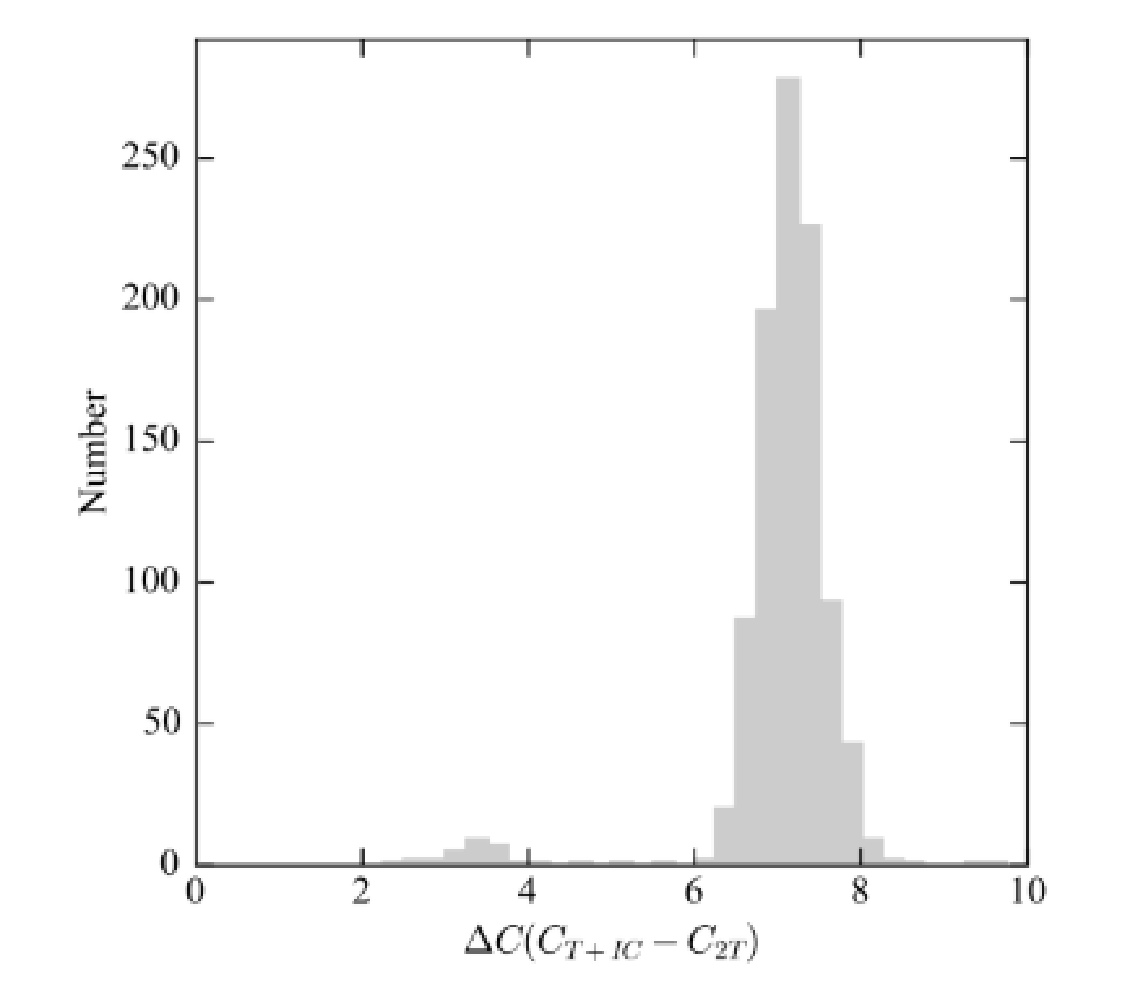}

\caption{The distribution of the difference in C-stat values ($\Delta$$C$) between the T$+$IC and 2T models from fits using the 1000 realizations of the background.  The 2T model is statistically better ($\sim$2.5$\sigma$ on average) at describing the \nustar-observed spectra than the T$+$IC model, with no realizations favoring it over the 2T model. Therefore, we conclude that the data clearly disfavor the addition of a non-thermal component.
}
%This histogram shows a 2.5$\sigma$ deviation of the 2T model from the T$+$IC model.} %replace with same image but with coordinates

\label{fig:statistics}

\end{figure}

    While this has shown that there is no significant IC scattering, it may still be the case that the T+IC model may fit better to the spectrum than the 2T model. To determine the model that most accurately describes the data given our knowledge of the background, we compare C-stat values from all fit iterations of the 2T and T$+$IC models to the global spectrum. 
We create a histogram of cash-statistic (C-stat) values stored from running 1000 iterations of each model (shown in Figure~\ref{fig:statistics}. The reason for doing so is that the magnitude of the C-stat depends on the number of bins used and the values of the data, so this does not inherently provide any information on the goodness-of-fit. This is remedied by repeatedly sampling new randomly generated data sets from the best-fit model, fitting them, and observing where the original C-stat is found in this distribution. These iterations are shown in Figure~\ref{fig:statistics}. 
    Based on Figure~\ref{fig:statistics}, we observe a ~2.5$\sigma$ deviation of the 2T model from the T$+$IC model. This suggests that the 2T model is the most appropriate model for characterizing the temperature distribution within this galaxy cluster. This makes sense, as the IC model has no exponential turnover, so the 2T model should fit the data better as long as there is no turnover in the data (which there isn't).
        When it comes to the 9T+IC model, the distribution of temperatures should theoretically be better captured. The reason why the 2T model is better fitting than this one has to do with the ARF generation. For the 2T model, we use the global region ARF shown in Figure~\ref{fig:region}, while in the 9T+IC model, each of the individual regions shown in Figure~\ref{fig:temperaturemap} has its own regional ARF, limiting the flexibility of the model to fit the data. This is also reflected in the difference in C-stat values between the two models. Our 9T+IC model has two free parameters (as discussed in Section~\ref{subsubsec:9T}) while our 2T model has 5 free parameters. That should give a difference in C-stat of 3, but the difference is greater (6), due to this lack of flexibility. 
    
    %Distribution of temperatures better captured in 9T, 2 free params vs 5 free params in 2T however, but C-stat difference is greater than 3. Global region arf is average for entire region, 9T arfs specific to individual regions, so not entirely capturing differences precisely. The shape
    
%\scalebox{0.7}{}

\section{Summary and Discussion} \label{sec:Summary}

\nustar\ observed A2163 for a time period of 115 ks which was then cleaned down to 112 ks after a manual filtering process. Prior to searching for a non-thermal signal, a detailed background emission model was applied to subtract background data from our spectra.

\subsection{Non-thermal Emission}

%9T+IC 1000 realizations, flux goes to 0 in x% of realizations.

    Based on our T$+$IC model, we can set a 90\% upper limit on the 20--80~keV flux of non-thermal emission coming from A2163 of $F_{\rm NT}$ $<$ 4.03 $\times$ $10^{-12}$~\flux\ using the T$+$IC model or $F_{\rm NT}$ $<$ 1.64 $\times$ $10^{-12}$~\flux\ using the 9T$+$IC model. As discussed in Section~\ref{subsubsec:best}, we can confidently rule out IC scattering when comparing the fits including a power law model to fits including two temperature models. The data favors a purely thermal model with two different components describing the variations in temperature across the cluster. It is important to note that the histogram in Figure~\ref{fig:histograms} still shows IC fluxes despite our claims that there is no significant detection. This presence is due to the IC flux masquerading as an extra temperature component. Our 9T+IC upper flux limit, when including systematic errors, goes nearly to 0, further suggesting that the IC interpretation is rejected. 
    %\todo{In this paragraph, you need to first discuss why the IC interpretation is rejected, especially because in table 2 in both the T$+$IC and 9T$+$IC fits, your IC flux is NOT consistent with 0!  In other words, someone will read that table and go, "but wait, their IC component detected with high significance -- at either 13.5 or 5.7 sigma!" (speaking of, are these uncertainties correct??).  Also, in Figure 8, none of your T$+$IC realizations have no IC flux, also indicating to the casual reader that IC is real.  You need to clearly explain why this is not the case.}
    The addition of an IC component does not describe the hard X-ray emission as well as the 2T model, at a confidence level of $\sim$2.5$\sigma$ based on Figure~\ref{fig:statistics}. Our non-thermal upper limit is an order of magnitude smaller than the limit obtained by \citet{2014A&A...562A..60O}, $F_{\rm NT} < 1.2 \times 10^{-11}$~\flux.
    Their constraint was limited by the sensitivity of the {\it Suzaku} HXD-PIN instrument. In comparison with the claimed detection from \RXTE, $F_{\rm NT} \sim 1.1^{+1.7}_{-0.7} \times 10^{-11}$~\flux, our limit is also an order of magnitude smaller. 
    The sensitivity of both instruments were limited by their substantially larger, non-imaging FOVs that admitted more cosmic background; the collimator designs admitted emission from $\sim$1\arcdeg~solid angles.
    %The accuracy of this detection is limited by \RXTE~being a collimator with a FWHM of 1 degree (the PSF looks like a triangle). 
    Their non-imaging nature means that the spatial origin of the detected emission is unknown;
    %we don't actually know where in the FOV the counts are coming from (
    the hardest emission could be coming from bright point sources unassociated with the ICM, for instance. 
    %{\it Suzaku} also suffers from this, although it is not as extreme. Local background information is also an issue with \RXTE. 
    
    \nustar's focusing optics allows point and other sources to be identified and excluded or avoided and provides less source-contaminated background measurements to be made concurrently during the observation, while {\it Suzaku} and \RXTE~did not. 
    On the other hand, the \beppo\ PDS, another non-imaging hard X-ray instrument, was able to monitor background conditions during observations by nodding between source and background fields.
    The instrument was still susceptible to nearby non-ICM sources, however.
    %\RXTE~also lacks the ability to nod, unlike \beppo~which has two collimators that nod which can provide more information on the background (which may vary with time). 
    If the cosmic background is not properly subtracted, hard emission from AGN could masquerade as a non-thermal signal. Our limit is more consistent with that from the \beppo~PDS ($F_{\rm NT} < 5.6 \times 10^{-12}$~\flux) \citep{2001A&A...373..106F}. %It is also comparable the limit obtained by \citet{2014ApJ...792...48W} for the Bullet Cluster ($F_{\rm NT} < 1.1 $\times$ 10^{-12}$~\flux).  \todo{What is the purpose of comparing your flux limit to that of the Bullet cluster?  Should they be comparable?  If so, explain why.} 

\subsection{Cluster Magnetic Field} \label{subsec:magnets}

    With an upper limit on the IC flux, we can set a lower limit on the average magnetic field strength $B$ using the ratio of the radio flux ($F_R$) to the X-ray flux ($F_X$) and the ratio of frequencies in the radio and X-ray bands where those fluxes are measured, $\nu_R$ and $\nu_X$, respectively. A total diffuse radio flux of 90~mJy inside our global extraction region was determined from VLA observations at 1.4~GHz. 
    %We determined the flux to be 90 mJy by using the size of the beams. 
    %Plugging that into the following equation:
    The magnetic field for a power law energy distribution of electrons emitting both synchrotron and IC emission can be determined from:

\begin{equation}
B = C(p)(1+z)^{(p+5)/(p+1)} \times \\
\bigg(\frac{F_R}{F_X}\bigg)^{2/(p+1)}
\bigg(\frac{\nu_R}{\nu_X}\bigg)^{(p-1)/(p+1)} \, ,
\end{equation}
where $p$ is the index of electron distribution ($N(E) \propto E^{-p}$ and related to the spectral index $\alpha$ by $p = 2\alpha + 1)$ and $C(p)$ is a proportionality constant %determined from the ratio of the synchrotron and IC fluxes
\citep{1979rpa..book.....R,1994hea..book.....L}.
This equation is just the extension of the relationship for one electron to a distribution of electrons at different energies and momenta. 
The ratio $F_R / F_X$ for a single electron is simply the ratio of the energy densities $U$ of the fields the electron is scattering:
\begin{equation}
\frac{F_R}{F_X} = 
\frac{U_B}{U_{\rm CMB}} = 
\frac{B^2/8\pi}{aT^4_{\rm CMB}} \, .
\end{equation}
%\todo{This is OK here, but consider moving this equation to the intro and just referring to it again here.}

With the T$+$IC model we obtain a lower limit of $B > 0.22~\mu$G, and with the 9T$+$IC model the limit is raised to 0.35~$\mu$G. These are both larger than previous limits quoted in \citet{2014A&A...562A..60O}, which is B $>$~0.098 $\mu$G for $\Gamma$~= 2.18 and $B > 0.006~\mu$G when using $\Gamma = 1.5$. 
%\todo{The feretti citation is confusing here -- is that just referring to the value of $\Gamma$?  Make clearer.  If that's the case, where does $\Gamma = 1.5$ come from?  Explain.}
The radio flux used in their work was 155~mJy at the same 1.4~GHz frequency. This varies from ours due to our smaller extraction region, which excluded roughly half the flux used in their study. % Ota et al. 2014
%\todo{Why does their radio flux differ from ours?  Did the estimate of the flux from the cluster change, or are we excluding close to half the flux using our smaller extraction region?  Explain here.}
Our limits both fall short of the $B = 0.4 \pm 0.2~\mu$G estimate found by \citet{2006ApJ...649..673R}. This limit is higher than ours due to different assumptions made about the distribution of relativistic electrons (as shown by the use of the spectral energy index $\Gamma = 1.6$ in their study). When using this value in our equation, our magnetic field limits increase by roughly 0.3~$\mu$G, consistent with the conflict in IC fluxes. 
%\todo{This explanation is too vague.  How does your limit on B change if you assume $\Gamma = 1.6$?  My guess would be you get a much higher lower limit on $B$ -- which would then be in conflict with the value they report but consistent with the conflict in IC fluxes.  You don't need to redo any fits, just plug in a different value for $p$ in equation 1.}
%Spectral index repeat calculation
%\todo{How is this estimate of $B$ so large if their flux measurement is 10$\times$ higher than our upper limit???}
Our lower limit is still short of the estimate for $B$ assuming equipartition conditions \citep[$B = 0.64~\mu$G;][]{2004A&A...423..111F}. These conditions assume that the total energy of a synchrotron source is distributed between fields and particles. An estimate for the total energy is taken at the minimum value. This condition is obtained by setting the magnetic field energy contributions equal to the contributions from the relativistic particles.
%which is the equipartition field. 
%\todo{Explain what the equipartition estimate is.}
Just like our non-thermal flux limit, our magnetic field most closely agrees with the lower limited estimated from \beppo\ %value obtained in \citet{2001A&A...373..106F} of 0.28 $\mu$G. %Rephaeli et al. 2006.
of 0.28~$\mu$G \citep{2001A&A...373..106F}.

\subsection{Temperature Map} \label{subsec:map}

While the focus of this paper is on the non-thermal emission present in the
cluster, we also briefly analyzed various regions of interest throughout the
cluster loosely based on the work done by \citet{2011AA...527A..21B}. %Bourdin et al. (2011). 
These regions and their temperatures are shown in Figure~\ref{fig:temperaturemap} and Table~\ref{table:t1}. Compared to the previously
referenced temperature map, created from \xmm\ data, we noticed the absence of high temperature ($\sim$18~keV) gas in the northeast of the cluster (region 9). 
A2163 lies somewhat near the Galactic plane in a region of high neutral hydrogen column density. 
If its distribution varies on small angular scales toward the cluster, but it is modeled as constant across the FOV, temperature estimates will be skewed by the fitting procedure,
which is weighted to focus the fit quality on the lower energy portion of spectra, where the count rate is higher. If this variation is taken into account, as was done in \citet{2011AA...527A..21B}, then another potential issue is model bias. In a future work (described in slightly more detail in the following section) we will be showing that depending on model selection (i.e. including molecular absorption and using different $n_{H}$ fitting models) the 18 keV gas can disappear entirely in the XMM-Newton.
\nustar's insensitivity to foreground absorption at the level present in A2163 removes this potential bias.  
%We also note a high temperature region ($\sim$15
%keV) concurrent with the other map \todo{what other map?}, as well as a cool core ($\sim$10 keV) located
%towards the center of the cluster, suggesting the movement of the merger to be
%along the northwesterly direction (while also pointing along the line of sight). 
%In future work, we will test the variable absorption hypothesis with joint fits of \nustar\ spectra with spatially coincident spectra from archival \xmm\ and \chandra\ observations.

\subsection{Future Work}

%\todo{I think it might make sense to move some of the above temperature map subsection to where you describe the 9T model, then discuss the absorption issue in this new section.  You can also move discussion of searching for IC emission spatially here, so the future work is all together in one place.}

In future work, we will revisit the topic of non-thermal emission locally within the cluster using \xmm~and \chandra~data to provide broad band, spatially resolved joint spectral fits with the \nustar~data. 
    %however due to uncertainties in the absorption and cross-correlation factors, we don't expect much improvement.
    If diffuse, non-thermal emission is more localized in the ICM, this approach could prove sensitive enough to detect it.
    However, due to uncertainties in the absorption and cross-correlation factors, we don't expect much improvement to the global constraint on IC emission from A2163.
        We will also revisit the temperature map presented in Figure~\ref{fig:temperaturemap} and test the variable absorption hypothesis discussed in Section~\ref{subsubsec:9T} with joint fits of \nustar\ spectra with the aforementioned spatially coincident spectra from archival \xmm\ and \chandra\ observations.

\acknowledgments
This work made use of data from the \nustar\ mission, a project led by the California Institute of Technology, managed by the Jet Propulsion Laboratory, and funded by NASA.
RARB and DRW gratefully acknowledge support from NASA grant 80NSSC19K0915. Basic research in radio astronomy at the Naval Research Laboratory is supported by 6.1 Base funding. 
This research has made use of the \nustar\ Data Analysis Software ({\tt NuSTARDAS}) jointly developed by the ASI Science Data Center (ASDC, Italy) and the California Institute of Technology (USA).
We thank the referee for useful comments.

\bibliography{NuSTARAbell2163}

%\noindent author year, title, version, publisher, prefix:identifier\\

%\citet{2015ApJ...805...23C} provides a example of how the citation in the
%article references the external code at
%\doi{10.5281/zenodo.15991}.  Unfortunately, bibtex does
%not have specific bibtex entries for these types of references so the
%``@misc'' type should be used.  The Repository tutorial explains how to
%code the ``@misc'' type correctly.  The most recent aasjournal.bst file,
%available with \aastex\ v6, will output bibtex ``@misc'' type properly.




\appendix

\section{ARF Generation} \label{app:ARF}

\subsection{Issue with {\tt numkarf}-generated ARFs}

During the analysis, we encountered a problem with the ARF generation routine {\tt numkarf} included in {\tt nuproducts}. 
The error is likely due to how the ARF is normalized using an unvignetted exposure map.
The issue arises when the {\tt extended=yes} flag is set, and it appears to be caused by something specific in this particular observation.
When the problem was initially discovered, a bug was found and fixed
within {\tt nuexpomap}, which caused the optical axis 
to be offset from its true location.
Unfortunately, that bug appears unrelated to this issue, or more
precisely, its fix did not fully solve the issue with ARF generation.

\begin{figure}[h]
\centering
\includegraphics[scale=0.75]{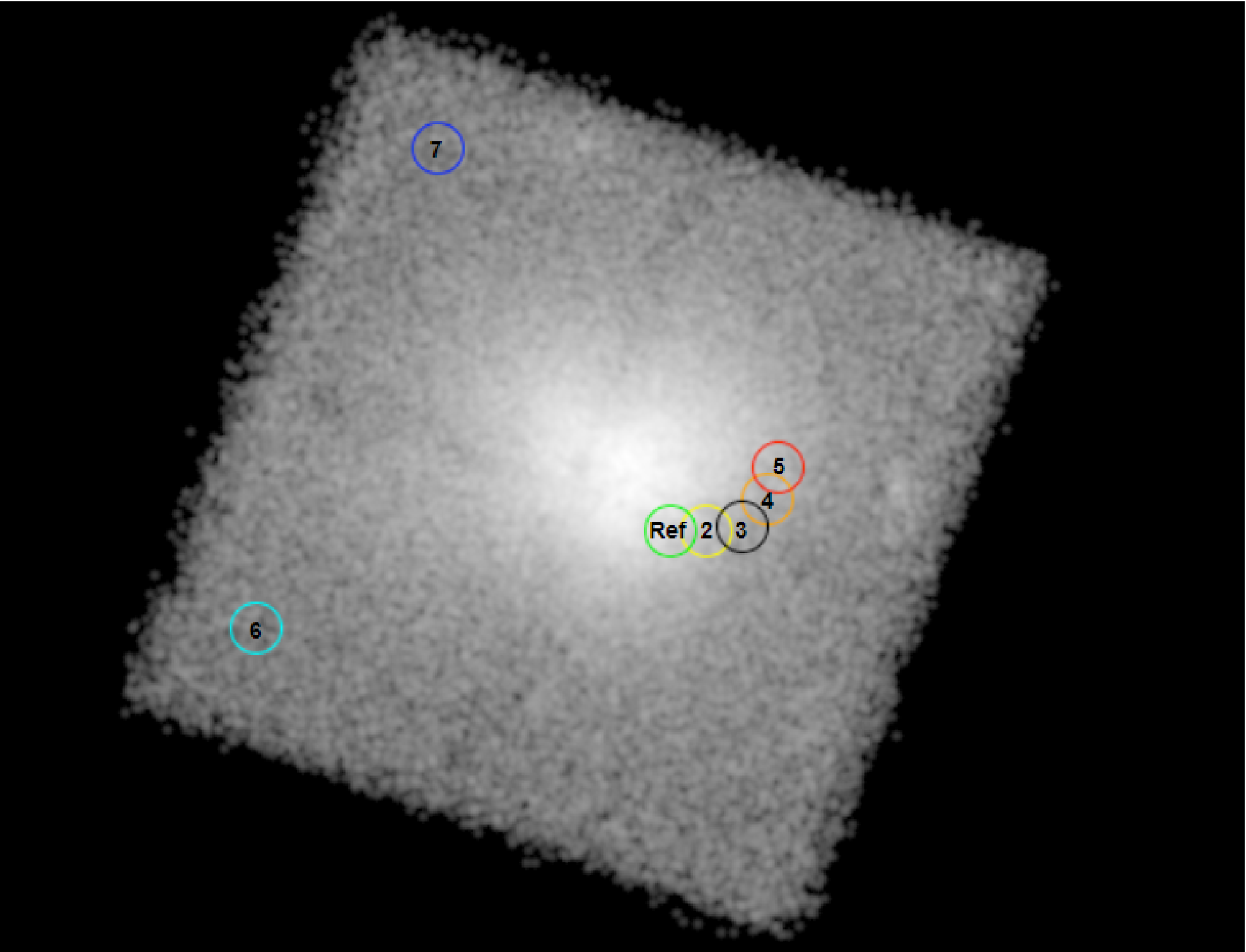}
\caption{The extraction regions used to check extended ARF generation with {\tt nuproducts}; their labels and colors correspond those used in Figure~\ref{fig:arfcheckplots}. 
%The regions are color coded to correspond with the legend in that figure. Regions 2 and 3 were used as checks. 
The reference region is placed near the location of the optical axis, which was determined using a vignetted exposure map at 10 keV.
%\todo{remake figs as outlined in email on 4/10/2020}
\label{fig:arfcheckimage}
}
\end{figure}

\begin{figure}[h]
\centering
\includegraphics[scale=0.625]{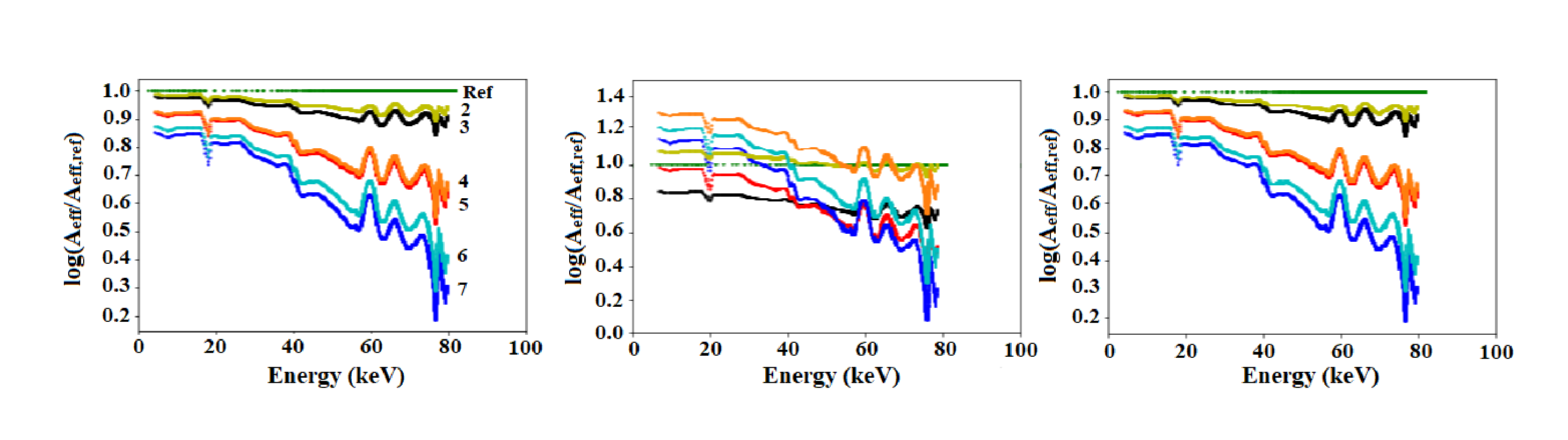}
%\includegraphics[scale=0.590]{test.PNG}
%\includegraphics[scale=0.585]{test2.PNG}
%\includegraphics[scale=0.525]{legend.png}
%\vspace{-4cm}
\caption{
Left: Point source ARFs, scaled by the Ref ARF (green line) representing the normalized ARF obtained from a region located near the optical axis. As expected, ARFs extracted from surrounding regions fall below the line, with effective area dropping more dramatically both in energy and in distance from the optical axis. 
Middle: ARFs generated using {\tt nuproducts} to generate extended ARFs from the same regions.
While the energy dependence of the ARF appears to be properly captured in each case, the overall normalizations are clearly incorrect.
%show that the optical axis is not being properly located, as ARFs generated from the same regions as before now lie above the normalized ARF. 
Right: ARFs generated using a custom script meant to apply the correct overall normalization to extended ARFs. 
All plots shown are from Telescope A. The same trend is seen in Telescope B. 
%\todo{remake figs as outlined in email on 4/10/2020}
\label{fig:arfcheckplots}
}
\end{figure}

To illustrate the issue, ARFs were created for several small circular
regions, shown in Figure~\ref{fig:arfcheckimage}.
Small region sizes were chosen so that the ARFs would be similar
for a region regardless of how they were created.
Both point source {\tt extended=no} and extended {\tt extended=yes} ARFs
were generated, with the latter set more appropriate for diffuse
ICM emission.
The region labeled ``Ref" is nearest the average position of the
optical axis; we plot all ARFs relative to this ARF in
Figure~\ref{fig:arfcheckplots}.
The point source ARFs (left panel) show the expected behavior, with farther
off-axis region ARFs having both lower overall normalization and
proportionately lower areas at higher energies, consistent with
the vignetting properties of \nustar.
While the extended ARFs (middle panel of Fig.~\ref{fig:arfcheckplots})
exhibit similar energy dependence as their corresponding 
point source ARFs, their overall normalizations do not.

\subsection{Custom ARF generation}

Point source ARFs are generated following the methodology employed
by {\tt numkarf}, albeit with independent code; test ARFs made
across the FOV differ by $<$1\% at all energies from those made
with {\tt numkarf}.
Extended ARFs are similarly produced as well in theory, although 
the code was written without reference to the corresponding code
in {\tt numkarf} to avoid recreating the issue.
First, point source ARFs are generated on a grid
over the region with spacings set by a {\tt boxsize} parameter
(default set to 10 pixels).
A weighted sum of these ARFs results in the final extended ARF.
The weighting normalizes each ARF to either the fraction of the
region area it covers (corresponding to a flat distribution of
source emission) or the fraction of counts within the region,
provided by a source image (generally taken to be a
background-subtracted \nustar\ ($E < 20$~keV) or other X-ray image.
In this implementation, chip gaps and excluded pixels are {\it not}
corrected for, but the impact of these corrections should be minimal. It should be noted that for this process, we used the non-exposure corrected image at first. Upon testing, however, the exposure corrected image and our original attempt agree to within 0.8$\%$ of each other.

The right panel of Figure~\ref{fig:arfcheckplots} shows the extended
ARFs made with this method.
While not identical to the {\tt numkarf}-generated point source
ARFs (as they shouldn't be, even for such small regions), the
ARFs are quite similar in both energy dependence and overall
normalization, demonstrating that they have been accurately derived.

\end{document}